\documentclass[showpacs,showkeys, groupedaddress,twocolumn, floatfix,aps, prb,reprint]{revtex4-1}
\usepackage[english]{babel}
\usepackage{fontenc}
\usepackage{graphicx}
\usepackage{amsmath}
\usepackage{epstopdf}
\usepackage{amssymb}
\usepackage{amsbsy}
\usepackage{amscd}
\usepackage{color}
\usepackage{bbm}
\usepackage{braket}
\usepackage{bm}
\usepackage{siunitx}
\usepackage[normalem]{ulem} 
\usepackage{verbatim}
\usepackage{url}
\usepackage[verbose,hypertexnames=false,bookmarksopenlevel=1,filecolor=blue,
linkcolor=blue,citecolor=blue,urlcolor=blue,pdfstartview=FitH,bookmarksopen,bookmarksnumbered,
colorlinks,plainpages=false,linktocpage]{hyperref}
\DeclareSymbolFont{mathbold}{OML}{cmm}{b}{it}
\DeclareMathSymbol{\bsigma}{\mathord}{mathbold}{27}
\begin{document}
\renewcommand{\figurename}{Fig.}
\title{Weak (anti)localization in tubular semiconductor nanowires with spin-orbit coupling}
\author{Michael Kammermeier}
\email{michael1.kammermeier@ur.de}
\author{Paul Wenk}
\author{John Schliemann}
\affiliation{Institute for Theoretical
  Physics, University of Regensburg, 93040 Regensburg, Germany}
\author{Sebastian Heedt}
\author{Thomas Sch\"apers}\affiliation{Peter Gr\"unberg Institute (PGI-9) and JARA-Fundamentals of Future Information Technology, Forschungszentrum J\"ulich, 52425 J\"ulich, Germany}
\date{\today }
\begin{abstract}
We compute analytically the weak (anti)localization correction to the Drude conductivity for electrons in tubular semiconductor systems of zinc blende type. 
We include  linear Rashba and Dresselhaus spin-orbit coupling (SOC) and compare wires of standard growth directions $\langle100\rangle$, $\langle111\rangle$, and $\langle110\rangle$. 
The motion on the quasi-two-dimensional surface is considered diffusive in both directions: transversal as well as along the cylinder axis. 
It is shown that Dresselhaus and Rashba SOC similarly
affect the spin relaxation rates. 
For the $\langle110\rangle$ growth direction, the long-lived spin states are of helical nature.
We detect a crossover from weak localization to weak anti-localization  depending on spin-orbit coupling strength as well as dephasing and scattering rate.
The theory is fitted to experimental data of an undoped $\braket{111}$ InAs nanowire device which exhibits a top-gate-controlled crossover from positive to negative magnetoconductivity.
Thereby, we extract transport parameters where we 
quantify the distinct types of SOC individually.
\end{abstract}
\pacs{71.70.Ej,72.25.Dc,72.25.-b,72.15.Rn,73.63.Hs,73.63.-b}
\keywords{semiconductor nanowire, weak localization, weak anti-localization, spin relaxation,  spintronics, spin-orbit coupling}
\maketitle
\section{Introduction}
 
In recent years semiconductor nanowires have gathered growing attention as they offer a large variety of applications such as lasers,\cite{Greytak2005} light-emitting diodes,\cite{Xing2015} photo-detectors,\cite{Xing2014} solar cells\cite{Krogstrup2013} and field-effect transistors\cite{Xiang2006} among others.
They, moreover, constitute an important platform in the search for Majorana bound states.\cite{Oreg2010,Mourik2012}
Catalytical growth using a "bottom-up" self-assembly technique from nanoparticles provides numerous possibilities to manipulate crystal structure, morphology and potential landscape.\cite{Fortuna2010}
Thereby, the nanowires, also often termed as nanorods or nanowiskers, are likewise highly interesting objects in the field of spintronics which exploits the spin degree of freedom of the electron in addition to its charge.\cite{Heedt2012}

Nanowires made from diamond and zinc blende type semiconductors tend to grow in the  $\braket{111}$ crystal direction as it minimizes the free energy.\cite{Fortuna2010}
However, the direction can be effectively controlled by substrate orientation, surface chemical treatment, temperature or pressure.\cite{Fortuna2010}
This also affects the nanowire's morphology/cross-sectional geometry\cite{Wacaser2006} and, remarkably, even the crystal structure.
It has been reported that nanowires, that are grown from materials which usually have zinc blende structure, are often polytypic with wurtzite segments\cite{Hiruma1995} or even exhibit pure wurtzite structure\cite{ZhangLu2014}. 

Another important feature is that axial or radial doping as well as the combination of different materials can change the potential landscape significantly.
With that, one is even able to design the transport topology of the current-carrying system.
Axial doping can generate \textit{pn} heterojunctions\cite{Haraguchi1992} or quantum dots\cite{Bjoerk2004,Hollosy2015}.
In narrow-gap semiconductors such as InAs, InSb, or InN due to Fermi level pinning the conduction band bends downwards at the surface of the nanowire and an electron accumulation layer is formed.\cite{Heedt2015,Hernandez2010,Bringer2011,BloemersDiss,Wirth2011}
However, using suitable dopants the potential profile can be flattened and the electrons uniformly distributed inside of the nanowire.\cite{Heedt2015,BloemersDiss,Wirth2011}
Moreover, the combination of different materials in core/shell nanowires confines the electrons either to a channel in the center or to a thin tubular layer a few nanometers below the surface of the nanowire.\cite{Lauhon2002,Bloemers2013}
In a different approach\cite{Haas2013} using etching techniques the core is removed and only the shell remains.

The huge degree of freedom in engineering those nanowires opens a vast amount of opportunities to study and manipulate spin-orbit interaction. 
Spin-orbit coupling (SOC) is the essential effect to control the spin and facilitate spintronic devices.
Depending on its origin one generally distinguishes between Rashba\cite{rashba_1} and Dresselhaus\cite{Dresselhaus1955b} SOC.
The latter results from different basis atoms in compound semiconductor materials and is sensitive to the crystal orientation and structure.
Rashba SOC occurs when an electric potential is present that exhibits an asymmetry. 
This can, for instance, either be induced externally by gating or internally by combinations of different semiconductor materials, doping or Fermi level pinning.
Both types of SOC lead to an effective magnetic field which is called spin-orbit field.

One prominent tool to study experimentally SOC are low-field magnetoconductance measurements. 
Quantum interference in disordered systems leads to a correction to the Drude conductivity.
The dimensionality of the system is of fundamental importance.
For low temperatures, if the SOC in a two-dimensional electron gas (2DEG) is weak or absent, the conductivity is reduced which is called weak localization (WL).
However, as a consequence of spin relaxation, the conductivity can be enhanced for strong SOC which is denoted weak anti-localization (WAL).
By applying a magnetic field the interference is destroyed as the time-reversal symmetry is broken.
Therefore, magnetoconductance measurements provide indirectly information about SOC.
By fitting the appropriate WAL theory to experimental data the spin relaxation time can be extracted which is related to strength and structure of the spin-orbit field. 
The theory of WAL has been developed for planar two-dimensional (2D) diffusive systems by Hikami \textit{et al.}, Ref.~\onlinecite{nagaoka}, and Iordanski \textit{et al.}, Ref.~\onlinecite{Iordanskii1994}.
The effect of hard-wall boundaries in quasi-one-dimensional planar wires was described by Kettemann, Ref.~\onlinecite{Kettemann2007a}, in the diffusive and by Kurdak \textit{et al.}, Ref.~\onlinecite{Kurdak1992}, in the ballistic regime.
Other works\cite{Skvortsov1998,Araki2014} analyzed the WL/WAL crossover  in the particular regime where the spin splitting becomes comparable or even exceeds the Bloch state uncertainty $\hbar/\tau$ due to the mean-free scattering time $\tau$.


It is highly topical among experimentalists to study WL/WAL in semiconductor nanowires.\cite{Hansen2005,Dhara2009,Roulleau2010,Liang2012,XiaoJie2010,
Hernandez2010,Weperen2015,Scheruebl2016}
Lacking a more precise theoretical description, 
many authors are compelled to apply existing theory even though it does not accurately match the system.
For instance, Refs.~\onlinecite{Hansen2005,Dhara2009,Roulleau2010,Liang2012,Scheruebl2016} investigate WL/WAL in $\braket{111}$ InAs nanowires by fitting magnetoconductance data with the formulas of Kurdak \textit{et al.}
However, in such systems the transport should be either governed by electron states at the surface due to Fermi level pinning or by states that are extended over the entire volume.\cite{Heedt2015,Hernandez2010,Bringer2011,BloemersDiss,Wirth2011} 
Both scenarios are not comprised in the theory of Kurdak \textit{et al.}
Also, it does not take into account the precise form of the spin-orbit field which has been proven to be significant in 2D systems.\cite{Knap1996,PhysRevB.90.115306,Wenk2015,Schliemann2003,Bernevig2006}
The spin relaxation rates due to Rashba and Dresselhaus SOC are not additive and the interplay of the fields can lead to a suppressed spin relaxation.\cite{Pikus1995,wenkdiss}
Additionally, the effects of Dresselhaus SOC are disregarded by all above-mentioned publications by referring to the vanishing spin splitting along  $\braket{111}$ in bulk zinc blende semiconductors.
Note that this argument would hold also for the $\braket{100}$ directions but it is widely known that in planar 2D systems the splitting is generally not absent along $\braket{100}$ owing to the structural confinement.\cite{winklerbook,Zutic2004a}
In nanowires, the situation can be similar.
Therefore, this statement applies only to quasi-3D wires.
Hence, the precision of the gained information about the system's transport parameters is limited.

In this paper, we develop an analytical model to describe the SOC effects in zinc blende nanowires with standard growth directions $\langle100\rangle$, 
$\langle111\rangle$, and $\langle110\rangle$.
The electrons are regarded to be radially confined to a thin surface layer where the cross-sectional geometry is approximated to be circular.
The motion on the quasi-2D surface of the cylinder
is considered diffusive in both directions: transversal as well as
along the cylinder axis. 
Including linear Rashba and Dresselhaus SOC we compute analytically the quantum mechanical correction to the Drude conductivity following the theory of Refs.~\onlinecite{Kettemann2007a, PhysRevB.83.115301, Wenk2010, wenkbook}. 
It is shown that
the Dresselhaus SOC will cause a shift of the triplet energy eigenmodes 
of the Cooperon Hamiltonian.  
In contrast to $\langle100\rangle$ and $\langle111\rangle$, the 
 low symmetry of the growth direction $\langle110\rangle$ generates an additional shift of the minimum of one triplet eigenmode to a finite value of the Cooperon momentum along the wire axis.
The gaps in the Cooperon spectrum are related to the spin relaxation rates.\cite{wenkdiss}
The relevant mechanism which is described by this theory is of D'yakonov Perel'\cite{perel}  type.
We detect a crossover from WL to WAL and from positive to negative
magnetoconductivity depending on SOC strength as
well as dephasing and scattering rate.  
A significant dependence on the wire width is not observed which is attributed to periodic boundary conditions along the circumference of the cylinder's surface.
The derived formulas serve as a model for zinc blende nanowires where a conductive tubular channel is formed either by Fermi level pinning or by structural confinement in core/shell nanowires.

This paper is structured as follows.
In the next section, we define the general bulk model for zinc blende semiconductors with SOC. 
We apply a coordinate transformation to a cylindrical system. 
In the next step, the quasi-2D surface model is developed including Rashba and Dresselhaus SOC where we restrict to terms linear in momentum.
Afterwards, we shortly discuss the conditions for spin conservation in such systems.
In Sec.~\ref{sec:QC}, we compute the Cooperon Hamiltonian for the 2D diffusive system including a magnetic field that is perpendicular to the wire axis.
We analyze the spectrum and derive approximations for its minima which are related to spin relaxation rates.
Finally, a formula for the correction to the Drude conductivity is derived.
In Sec.~\ref{sec:fit}, we fit the derived formulas to magnetoconductance measurements of a top-gated $\braket{111}$ InAs nanowire.
We recover the gate-controlled crossover from positive to negative magnetoconductivity and gather information about spin relaxation and dephasing rates as well as SOC strengths and radial confinement.


%


\allowdisplaybreaks

%
\section{Model Hamiltonian}
\subsection{Bulk Model}
Throughout this work we set $\hbar=1$. We start with the bulk
Hamiltonian $\mathcal{H}$ for electrons in the $\Gamma_{6c}$
conduction band with SOC as
\begin{align}
\mathcal{H}&=\,\frac{\mathbf{k}^2}{2 m}+\mathcal{H}_\text{R}+\mathcal{H}_\text{D}^{[001]}.
\label{nanorodbulk}
\end{align}
The terms
\begin{align}
\mathcal{H}_\text{R}&=r_{41}^{6c6c}\left[(k_y \mathcal{E}_z - k_z \mathcal{E}_y)\sigma_x+\text{c.p.}\right],\label{rashba}\\
\mathcal{H}_\text{D}^{[001]}&=b_{41}^{6c6c}\left[\{k_x,k_y^2-k_z^2\}\sigma_x+\text{c.p.}\right], \label{dresselhaus}
\end{align}
denote the Rashba $\mathcal{H}_\text{R}$ and Dresselhaus
$\mathcal{H}_\text{D}$ SOC contributions with the material specific
parameters $r_{41}^{6c6c}$ and $b_{41}^{6c6c}$, $\mathcal{E}_i$ the
electric field components, $\sigma_i$ the Pauli matrices, $m$ the
effective electron mass and $\{A,B\}=(AB+BA)/2$ is the symmetrized
anticommutator.\cite{winklerbook} 

In this definition, the basis vectors correspond to the
$\left\langle100\right\rangle$ crystal axes.   
As we also consider $\left\langle 111\right\rangle$ and $\left\langle 110\right\rangle$ nanowires, we rotate the Hamiltonian
such that the new basis vectors are aligned with the new crystal axes.
In general, we define the $z$ axis to be parallel the wire's growth direction.
The rotation can be
performed by means of the rotation operator $\mathcal{D}$ which
transforms an arbitrary vector $\mathbf{v}$ as
\begin{align}
\mathbf{v}\mapsto \mathcal{D}\, \mathbf{v}.
\label{vecrot}
\end{align}
The rotation operator $\mathcal{D}$ is given by
\begin{align}
\mathcal{D}(\theta,\phi)&=
\begin{pmatrix}
\cos(\phi)\cos(\theta)&-\sin(\phi)&\cos(\phi)\sin(\theta)\\
\sin(\phi)\cos(\theta)&\cos(\phi)&\sin(\phi)\sin(\theta)\\
-\sin(\theta)&0&\cos(\theta)	
\end{pmatrix},
\label{rotmatrix}
\end{align}%
where $\theta$ denotes the polar and $\phi$ the azimuth angle of the
former coordinate system, that is,
$\theta=\arccos\left(1/\sqrt{3}\right)$ and $\phi=\pi/4$ for $\left\langle 111\right\rangle$ nanowires and $\theta=\pi/2$ and $\phi=\pi/4$ for $\left\langle 110\right\rangle$ nanowires.  
An additional rotation about the transformed $z$ axis can be applied to choose the alignment of the $x$ and $y$ basis vectors with the crystallographic axes of the new system as desired.
Here, we select for $\left\langle 111\right\rangle$ nanowires the Cartesian basis system as
 $\hat{x}\parallel[11\overline{2}]$, $\hat{y} \parallel
[\overline{1}10]$, $\hat{z}\parallel[111]$ and for $\left\langle 110\right\rangle$ nanowires as 
$\hat{x}\parallel[\overline{1}10]$, $\hat{y} \parallel
[001]$, $\hat{z}\parallel[110]$.
The Rashba Hamiltonian is invariant with respect to
rotation of the crystal provided that the electric field rotates
analogously.  However, in confined systems as shown for a
2DEG\cite{Zutic2004a} the Dresselhaus Hamiltonian depends on the
crystal orientation.  In the transformed coordinate systems, it takes the form
\begin{align}
\mathcal{H}_\text{D}^{[111]}=&\,\frac{b_{41}^{6c6c}}{2\sqrt{3}}\left\{\left[-k_y\left(k_x^2+k_y^2+2\sqrt{2}k_x k_z-4k_z^2\right)\right]\sigma_x\right.\notag\\
&\,+\left[k_y^2\left(k_x+\sqrt{2}k_z\right)\right.\notag\\
&\quad\;\left.+k_x\left(k_x^2-\sqrt{2}k_x k_z-4 k_z^2\right)\right]\sigma_y\notag\\
&\,\left.+\left[\sqrt{2}k_y\left(3k_x^2-k_y^2\right)\right]\sigma_z\right\}
\label{HD111}
\end{align}
and
\begin{align}
\mathcal{H}_\text{D}^{[110]}=&\,b_{41}^{6c6c}\left\{\frac{1}{2}k_z\left(k_x^2+2k_y^2-k_z^2\right)\sigma_x\right.\notag\\
&\,-2 k_x k_y k_z\sigma_y\notag\\
&\left.\,+\frac{1}{2}k_x\left(-k_x^2+2k_y^2+k_z^2\right)\sigma_z\right\}.
\label{HD110}
\end{align}
Semiconductor nanowires often exhibit a cross-sectional geometry of a hexagon.\cite{Fortuna2010} 
Nevertheless, for simplicity we will assume the nanowire
to have cylindrical symmetry in the following.
Thus, we introduce cylindrical coordinates.

\subsection{Coordinate Transformation}
The Cartesian and the cylindrical coordinates are related through the
equations
\begin{align}
r=&\,\sqrt{x^2+y^2},\\
\phi=&\,\arctan\left(\frac{y}{x}\right),
\label{CartCyl}
\end{align}
where the inverse tangent is suitably defined to take the correct quadrant of $(x,y)$ into account.
Hence, in a cylindrical system the wave vector
$\mathbf{k}=(k_x,k_y,k_z)^\top$ and the vector of Pauli matrices
$\boldsymbol{\sigma}=(\sigma_x,\sigma_y,\sigma_z)^\top$ transform into
\begin{align}
\mathbf{k}=&\,\mathbf{\hat{r}}\,k_r+\boldsymbol{\hat{\phi}}\,k_\phi+\mathbf{\hat{z}}\,k_z,\\
\boldsymbol{\sigma}=&\,\mathbf{\hat{r}}\,\sigma_r+\boldsymbol{\hat{\phi}}\,\sigma_\phi+\mathbf{\hat{z}}\,\sigma_z,
\label{cylvec}
\end{align}
where $k_r=-i \partial_r$, $k_\phi=- \frac{i}{r} \partial_\phi$,
$k_z=-i \partial_z$. 
The orthonormal unit vectors in the Cartesian basis are
\begin{align}
\mathbf{\hat{r}}=
\begin{pmatrix}
\cos(\phi)\\
\sin(\phi)\\
0\\
\end{pmatrix},
\,\boldsymbol{\hat{\phi}}=
\begin{pmatrix}
-\sin(\phi)\\
\cos(\phi)\\
0\\
\end{pmatrix},
\,\mathbf{\hat{z}}=
\begin{pmatrix}
0\\
0\\
1\\
\end{pmatrix}.
\label{unitveccyl}
\end{align}
Therefore, the time-independent Schr\"odinger equation for $\mathcal{H}$ becomes
\begin{align}
\bigg[&-\frac{1}{2m}\left(\partial_r^2+\frac{1}{r}\partial_r+\frac{1}{r^2}\partial_\phi^2+\partial_z^2\right)\notag\\
&+V(\mathbf{r})+\mathcal{H}_\text{R}+\mathcal{H}_\text{D}\bigg]\ket{\psi}=\,E \ket{\psi},
\label{H0cyl}
\end{align}
where we included a position-dependent potential $V(\mathbf{r})$ that causes a structural confinement to be discussed in the following subsection.  
We identify in the component $k_\phi$ the angular momentum operator along the $z$ axis: $\mathcal{L}_z=-i \partial_\phi$.  The transformation into the cylindrical coordinate system has an important consequence.  
In the new Hamiltonian position operators $r,\phi$ occur and one has to take
account of the non-commutativity with the momentum operators
$k_r,k_\phi$.  
These position operators are also implicitly contained
in $k_\phi$, $k_r$, $\sigma_r$, and $\sigma_\phi$.  Yet, since
$[k_r,\sigma_i]=[k_\phi,\sigma_i]=\,0$ where $i \in \{x,y,z\}$, it is often convenient to keep the Pauli matrices Cartesian.  
The Pauli matrices in cylindrical coordinates and the relevant commutation relations are given in the Apps. \ref{app:pauli} and \ref{app:com}.
 Owing to these commutators, the Hermiticity  of the derived model Hamiltonian is often not obvious.\cite{Aronov1993}

\subsection{Tubular System}\label{subsec:tub}
Hereafter, we follow the procedure used to derive a quasi-one-dimensional Hermitian Hamilton operator for mesoscopic rings in
presence of SOC as done in Refs.~\onlinecite{Meijer2002,Sheng2006,Berche2010}.

%
\begin{figure}[t]
\includegraphics[width=.85\columnwidth]{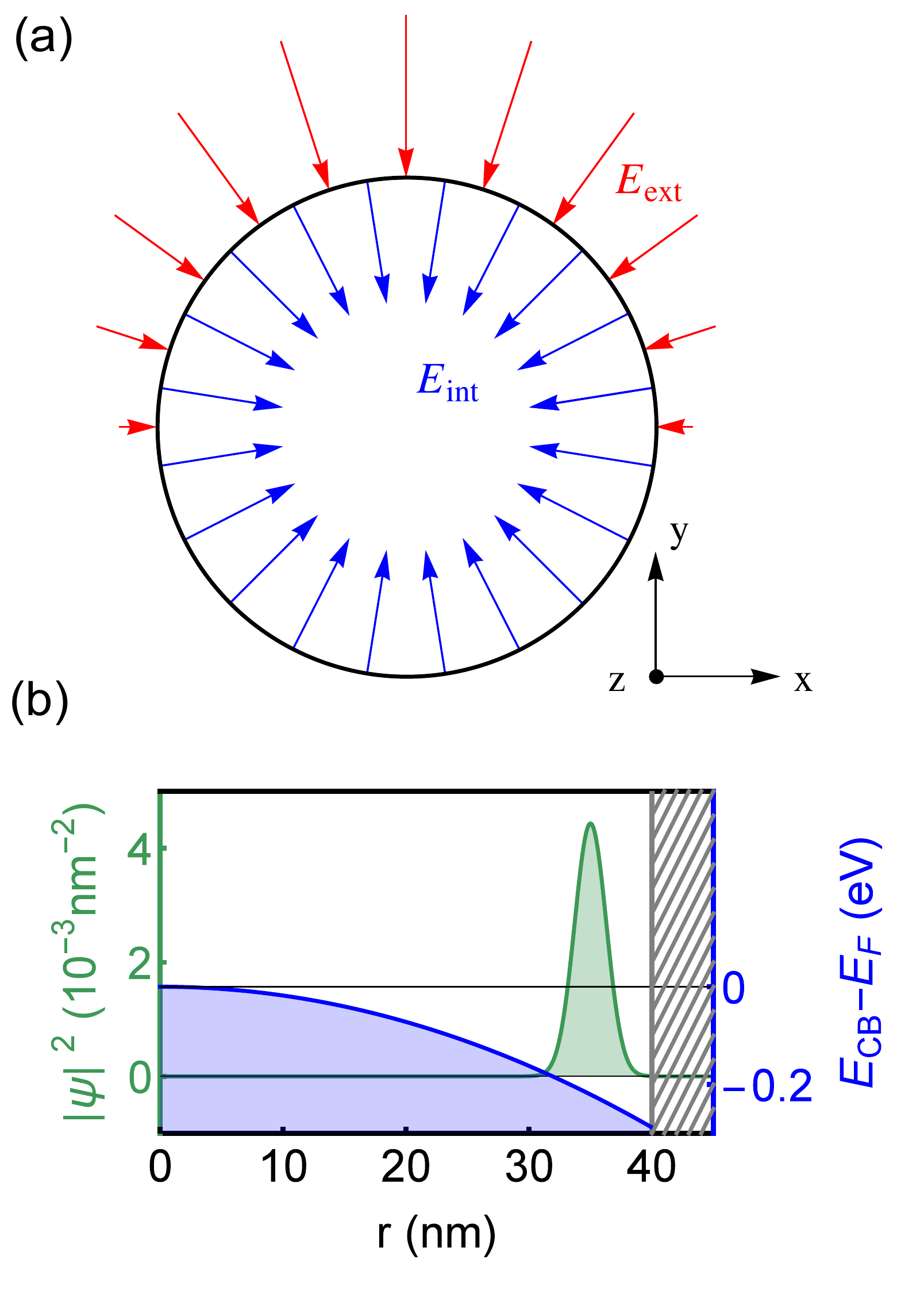}
\caption{(Color online) (a) Internal (blue) and external (red) electric field that lead to Rashba SOC in a nanowire (here, $E_\text{int},E_\text{ext}<0$). 
The internal field can be a consequence of Fermi level pinning, the external due to a gate voltage.
Fig. (b) sketches the situation of a nanowire with radius $R_0=\SI{40}{nm}$.
Here, the electron probabitity density $|\psi|^2$ (green) is focused at $R=\SI{35}{nm}$ below the surface and extends over an area of about $\SI{10}{nm}$ for a confinement parameter $\gamma=\SI{0.55}{nm^{-1}}$. 
The blue line illustrates the bending of the conduction band (CB) edge due to Fermi (F) level pinning.
The resulting radial confinement is modelled by a harmonic potential in this work.
 }
\label{pic:wavefct_plot}
\end{figure}

In order to obtain the tubular geometry of the nanowire, we consider a
radial harmonic confinement potential
$V(\mathbf{r})=V(r)=\frac{1}{2}m\omega^2(r-R)^2$ which forces the
electron wave function to be localized at a narrow region around the
cylinder radius $R$. \footnote{Utilization of a harmonic potential is
  particularly convenient since most of the matrix elements take a
  simple form.}  $R$ is assumed to be large in comparison with the mean
radial extent of the wave function.  If the potential is
steep enough, the particles fill the lowest radial eigenmode only.
Hence, we can treat the Hamiltonian perturbatively by separating
\begin{align}
\mathcal{H}=\,\mathcal{H}_0+\mathcal{H}_1,
\end{align}
where
\begin{align}
\mathcal{H}_0=&\,-\frac{1}{2 m}\left(\partial_r^2+\frac{1}{r}\partial_r\right)+\frac{1}{2}m\omega^2(r-R)^2,\label{h0hollow}\\
\mathcal{H}_1=&\,-\frac{1}{2 m}\left(\frac{1}{r^2}\partial_\phi^2+\partial_z^2\right)+\mathcal{H}_\text{R}+\mathcal{H}_\text{D}.\label{h1hollow}
\end{align}
In the limit of a 2D tubular system we can neglect the term $\frac{1}{r}\partial_r$ in comparison with $\partial_r^2$. 
Thus, the Schr\"odinger equation for $\mathcal{H}_0$ reduces to a one-dimensional harmonic oscillator equation.
The normalized eigenfunction for the lowest radial mode is given by
\begin{align}
\braket{r|R_0}=\left(\frac{\gamma}{\sqrt{\pi}R}\right)^{1/2}\exp\left[-\frac{\gamma^2}{2}(r-R)^2\right],
\label{f(r)eigenfkt}
\end{align}
where $\gamma^2=m\omega$ and the ground state eigenenergy
$E_{R_0}=\omega/2$.  The 2D approximation is justified
since
\begin{align}
\frac{\braket{\frac{1}{r}\partial_r}}{\braket{\partial_r^2}}\approx-\frac{e^{-\gamma ^2 R^2}}{\sqrt{\pi } \gamma  R}\stackrel{\gamma R\rightarrow \infty}{\longrightarrow} 0 
\label{small}
\end{align}
for $\gamma R\gg 1$. Note that in order to obtain
analytical expressions the integrals have to be extended to
$r\,\epsilon\, [-\infty,\infty]$.  This, however, is applicable since
we assume $\braket{r|R_0}\approx0$ for $r\leq0$.\footnote{For $R
  \gamma>2$ the deviation $\Delta =|\braket{R_0|R_0}-1|\propto
  10^{-4}$ and
  $\braket{\frac{1}{r}\partial_r}/\braket{\partial_r^2}\propto10^{-3}$.} 
In this approximation, Eq.~(\ref{small}) vanishes exactly. 

The quasi-2D tubular Hamiltonian is now defined as
\begin{equation}
\mathcal{H}_\text{tube}\equiv\,\braket{R_0|\mathcal{H}_1|R_0}.
\label{tubeHam}
\end{equation}
Making use of the fact that $\gamma R \gg 1$ and the SOC terms are assumed to be small compared to the kinetic part of the Hamiltonian, we keep only terms of the order of $\mathcal{O}(1/r)$ in the Dresselhaus Hamiltonian.  
The remaining relevant matrix elements are given in App. \ref{app:me}.  We stress that in contrast to a 2DEG the matrix elements for the momentum
operator along the confinement direction $\braket{k_r^3}$ and
$\braket{k_r}$ do not vanish.
The latter was disregarded in Ref.~\onlinecite{Spisak2014} by 
parity arguments which do not hold for a cylindrical system.
In fact it is possible to show that independent of the exact form of $V(\mathbf{r})$ one obtains $\braket{k_r}=i/(2 R)$ which we prove in App.~\ref{app:kr}.
Since the model results from $\mathbf{k}\cdot\mathbf{p}$ theory, using an expansion of $\mathbf{k}$ around the $\Gamma$ point, we can also neglect the subordinate terms $\propto k_z^2$ and
obtain a fully linearized version of the Dresselhaus SOC:
%
\begin{align}
\mathcal{H}_\text{D}^\text{[001],2D}=&\,\beta
\left\{\sigma_r\left[\frac{1}{2}\sin(2\phi)k_\phi-2\cos(2\phi)\braket{k_r}\right]\right.\notag\\
&\,-\sigma_\phi\left[\cos(2\phi)k_\phi+\frac{5}{2}\sin(2\phi)\braket{k_r}\right]\notag\\
&\,+\sigma_z \cos(2\phi)k_z\Big\},\label{HDlin001}\\
\mathcal{H}_\text{D}^\text{[111],2D}=&\,\frac{\beta}{2\sqrt{3}}\left\{\sigma_r\left[-\sqrt{2}\sin(3\phi)k_z-k_\phi\right]\right.\notag\\
&\,+\sigma_\phi\left[\braket{k_r}-\sqrt{2}\cos(3\phi) k_z\right]\notag\\
&\,\left.+\sigma_z\, 3 \sqrt{2}\Big[\cos(3 \phi) k_\phi+ 3 \sin(3\phi)\braket{k_r}\Big]\right\},\label{HDlin111}\\
\mathcal{H}_\text{D}^\text{[110],2D}=&\,\frac{\beta}{8}\left\{\sigma_r\, \Big[\cos(\phi)+3\cos(3\phi)\Big]k_z\right.\notag\\
&\,-\sigma_\phi\,\Big[11\sin(\phi)+3\sin(3\phi)\Big] k_z\notag\\
&\,+\sigma_z 
\left\{\Big[\sin(\phi)+9\sin(3\phi)\Big]k_\phi\right.\notag\\
&\left.\left.\phantom{+\sigma_z \Big\{ }-\Big[\cos(\phi)+27\cos(3\phi)\Big]\braket{k_r}
\right\}\right\},\label{HDlin110}
\end{align}
with $\beta=b_{41}^{6c6c}\braket{k_r^2}$.
Quite recently, the tubular Dresselhaus Hamiltonian for the $[111]$ growth direction has also been derived by Kokurin in Ref.~\onlinecite{Kokurin2015} in a similar way, using a different alignment of the $x$ and $y$ axes.

Concerning the Rashba SOC, we can distinguish two different sources
for an electric field.  
First, similarly to the case of a planar 2DEG, we assume a constant and homogeneous internal electric field pointing in the direction of the confinement, i.e., $\boldsymbol{\mathcal{E}}_\text{int}=\mathcal{E}_\text{int}\,\mathbf{\hat{r}}$ with $\mathcal{E}_\text{int}$ being constant.
This field is a consequence of Fermi level pinning which can be altered by doping.  
Second, since the Rashba effect can be modified externally by a gate voltage, we simulate a realistic situation for the experiment as the one performed by Heedt \textit{et al.}, Ref.~\onlinecite{Heedt2015}.
There, the gate electrode is laterally fixed (in this model chosen to be in the
$\mathbf{\hat{y}}$ direction) to the wire leading to an inhomogeneous
field.
We approximate it by
$\boldsymbol{\mathcal{E}}_\text{ext}=\,\mathcal{E}_\text{ext}\sin
(\phi) \Theta(\phi)\Theta(\pi-\phi)\,\mathbf{\hat{r}}$ where $\Theta$ is the Heaviside function.  
Both fields and the resulting radial confinement for the wave function are schematically depicted in Fig.~\ref{pic:wavefct_plot}.
Since the field $\boldsymbol{\mathcal{E}}_\text{ext}$ depends on the polar angle $\phi$ that does not commute with $k_\phi$, we need to symmetrize the Rashba Hamiltonian in
Eq.~(\ref{rashba}) in order to obtain a Hermitian operator.
Consequently, we find for the Rashba SOC contribution
\begin{align}
\mathcal{H}_\text{R}^\text{2D}=&\,\alpha_\text{int} \Big[\sigma_\phi k_z -\sigma_z  k_\phi  \Big]\notag\\
&+\alpha_\text{ext} \Theta(\phi)\Theta(\pi-\phi)\Big\{\sigma_\phi \sin(\phi) k_z\notag\\
&\phantom{+}+\sigma_z \Big[\cos(\phi) \braket{k_r}-\sin(\phi )k_\phi\Big]\Big\},
\label{rashbacyl}
\end{align}
with $\alpha_\text{int/ext}=r_{41}^{6c6c} \mathcal{E}_\text{int/ext}$.
In  Eqs.~(\ref{HDlin001})-(\ref{rashbacyl}) the order of the operators $\phi$, $k_\phi$, $\sigma_r$, and $\sigma_\phi$ is crucial.
The internal Rashba Hamiltonian has been set up previously\cite{Trushin2007,Bringer2011} to study spin dynamics in cylindrical 2DEGs and similarly in curved 1D wires\cite{Trushin2006}.

At this point, we emphasize that this derivation is fundamentally
different from a previous model considered by Magarill \textit{et
  al.}, Ref.~\onlinecite{Magarill1998}, and Manolescu \textit{et al.}, Ref.~\onlinecite{Manulescu2013}. 
These authors used the Rashba and linearized Dresselhaus Hamiltonian of a $[001]$ confined 2DEG and transformed the in-plane Cartesian coordinates into the in-plane coordinates of the cylinder surface.
In other words, they described a 2DEG wrapped around a core to form the shell of a hollow cylinder.  
In case of the Rashba contribution (referring to the internal part of
$\mathcal{H}_\text{R}^\text{2D}$) both situations do not differ.
The reason is that the intrinsic electric field is equivalent as in both cases it is assumed to penetrate the surface perpendicularly, i.e., the field is collinear with the confinement direction. 
However, as in both scenarios the structure of the crystal in the layer is different, the Dresselhaus contribution will be distinct as well.  
The model of Refs.~\onlinecite{Magarill1998} and \onlinecite{Manulescu2013} requires a deformation of the crystal structure.  
Thus, if the radius $R$ of this cylinder is small the effects due to strain are presumably very important.
On the other hand, if the radius $R$ of the cylinder is large the
situation becomes nearly equivalent to a flat 2D system with a periodic
boundary condition for one of the in-plane vectors.  
The approach used in this publication does not assume a deformed crystal and strain effects are less relevant.
Moreover, it was shown in Ref.~\onlinecite{Haas2013} that the crystal structure in the shell of a core/shell nanowire can adopt the structure of the core.
Therefore, a rolled-up 2DEG seems not to be the proper model for a realistic core/shell nanowire.

Returning to our model, we can express the quasi-2D Hamilton operator $\mathcal{H}_\text{tube}$ as a matrix with the normalized basis functions 
\begin{align}
\braket{\phi|\,l}=\,\frac{1}{\sqrt{2\pi}}\exp(i l \phi),\label{eigenfcthollowH0phi}\\
\braket{z|\,k_z}=\,\frac{1}{\sqrt{L}}\exp(i k_z z),\label{eigenfcthollowH0z}
\end{align}
where $l$ is the angular momentum quantum number.  Moreover, we assume
periodic boundary conditions in axial direction with periodicity $L$
leading to plane wave solutions with the quasi-continuous quantum
number $k_z$.
It is worth to mention that in a system with only internal Rashba SOC the Hamiltonian commutes with the $z$-component of the total angular momentum operator, i.e., $\mathcal{J}_z=\mathcal{L}_z+\frac{1}{2}\sigma_z$. \cite{Bringer2011} 
Yet, as soon as external Rashba or Dresselhaus SOC are incorporated the total angular momentum $j=l\pm1/2$ is no more a good quantum number.

\subsection{Spin Conservation on the Tubular Surface}

Commonly, in systems with SOC the spin rotation symmetry is broken.
As the spin precession depends on the momentum of the carrier, scattering in a diffusive semiconductor with inversion asymmetry randomizes the spin which results in D'yakonov Perel'\cite{perel} spin relaxation.
In planar 2D electron or hole systems, however, the combination of Rashba and Dresselhaus SOC and perhaps strain effects can lead to 
an SU(2) spin rotation symmetry which is robust against spin-independent disorder as demonstrated in
Refs.~\onlinecite{Schliemann2003,Bernevig2006,PhysRevB.90.115306,Wenk2015}.
Similarly, Trushin \textit{et al.}\cite{Trushin2007} showed that in a rolled-up 2DEG a certain ratio of Rashba SOC strength and curvature radius leads to a conservation of the tangential spin component $\sigma_\phi/2$.

In the tubular nanowires studied in this article we assume diffusive motion and treat the transverse momentum $k_\phi$ in the same way as $k_z$, as a quasi-continuous quantity.
We will average over all in-plane momenta and azimuthal angles which will become clear in the subsequent section.
As a result, the interplay between Rashba and Dresselhaus SOC does not lead to a suppressed D'yakonov Perel' spin relaxation in a tubular nanowire grown along the high-symmetry directions $\braket{001}$ or $\braket{111}$.
We can refer this characteristic to the fact that in general Rashba and Dresselhaus SOC exhibit a different $\phi$ dependence.
The mismatch is particulary pronounced for the internal Rashba contribution as it is independent of $\phi$.
Thus, the interplay between Rashba and Dresselhaus cannot generate a collinear field independent of its azimuthal location. 
On the other hand, for $\braket{110}$ nanowires due to their lower symmetry we will observe that the internal Rashba as well as the Dresselhaus SOC  compete with the external Rashba SOC. 
Therefore, the ratios of interaction strengths modify the spin relaxation rate.
Moreover, we will find that both Rashba and Dresselhaus SOC yield an additional shift of the Cooperon triplet spectra that cause an insuppressible spin relaxation.
It is also to mention that the particular scenario found by Trushin \textit{et al.}\cite{Trushin2007} is not reflected in our results for the Cooperon spectrum.
We attribute this property to the azimuthal averaging which diminishes the curvature effects. 

\section{Quantum Correction to the  Conductivity}\label{sec:QC}
By means of diagrammatic perturbation theory, we can construct the first-order
correction to the Drude conductivity, $\Delta\sigma$, which results from
quantum interference between self-crossing paths in a disordered conductor.  We assume the
following preconditions on the impurity potential
$V_\text{imp}(\mathbf{r})$: First, we consider a standard white-noise
model for the impurity potential, that is, it vanishes on average
$\left\langle V_\text{imp}(\mathbf{r})\right\rangle$ and is
uncorrelated, i.e., $\left\langle V_\text{imp}(\mathbf{r})V_\text{imp}(\mathbf{r'})\right\rangle\propto\delta(\mathbf{r}-\mathbf{r'})$.
Second, we assume weak disorder, i.e., $\epsilon_F \tau \gg 1$, where $\epsilon_F$ is the
Fermi energy and $\tau$ is the mean elastic scattering time.  We also
consider the electrons' motion to be diffusive in longitudinal as well
as transversal direction of the 2D cylindrical shell.  By averaging
over all impurities and summing up all maximally crossed ladder
diagrams, we find the quantum correction to the longitudinal static
conductivity~\cite{nagaoka} to first order in $\left (\epsilon_F \tau\right )^{-1}$ given by the real part of the
Kubo-Greenwood formula
\begin{align}
\Delta\sigma=&\,-\frac{e^2 }{\pi}\frac{D_e}{ \mathcal{V}}\times\notag\\
&\Re \text{e}\left(\sum_{\mathbf{Q}}\sum_{s_1,s_2=\pm}\braket{\mathbf{Q}|\braket{s_1,s_2|\hat{\mathcal{C}}(\mathbf{\hat{Q}})|s_2,s_1}|\mathbf{Q}}\right).
\label{conductivity}
\end{align}
Here, $\mathcal{V}$ is the surface of the nanowire, $D_e$ the
diffusion constant in two dimensions, i.e., $D_e=v_F^2 \tau/2$, with
the Fermi velocity $v_F$, $s_i$ the spin-$z$ quantum number,
$\hat{\mathcal{C}}$ the Cooperon propagator, and
$\mathbf{Q}=\mathbf{p}+\mathbf{p'}$ the sum of the electron momentum
and the electron's momentum along its time-reversed path.  
Below, we follow the approach in Refs.~\onlinecite{Kettemann2007a, PhysRevB.83.115301, Wenk2010, wenkbook} to
compute the quantum correction to the conductivity.
%
\subsection{Cooperon Hamiltonian}
The Cooperon propagator $\hat{\mathcal{C}}$ for 
low temperature can be approximated by
\begin{equation}
\hat{\mathcal{C}}(\mathbf{\hat{Q}})=\,\tau\left(1-\hat{I}(
\mathbf{\hat{Q}})\right)^{-1}
\label{cooperon1}
\end{equation}
with the correlation function
\begin{align}
\hat{I}(\mathbf{\hat{Q}})={}&\frac{1}{2\pi\nu\tau\mathcal{V}}\sum_\mathbf{q}\braket{\mathbf{q}|\,\mathcal{G}^\text{R}(\mathbf{\hat{q}},\boldsymbol{\sigma})\mathcal{G}^\text{A}(\mathbf{\hat{Q}}-\mathbf{\hat{q}},\boldsymbol{\sigma'})\,|\mathbf{q}},
\label{correl}
\end{align}
where $\nu=m/(2\pi)$ is the 2D density of states per spin. 
The retarded/advanced Green's operator $\mathcal{G}^\text{R/A}$ for positive energy yields in first-order Born approximation
\begin{equation}
\mathcal{G}^\text{R/A}(\mathbf{q},\boldsymbol{\sigma})=\,\frac{1}{\epsilon_F-\mathcal{H}(\mathbf{q},\boldsymbol{\sigma})\pm \frac{i}{2\tau}}
\label{GF}
\end{equation}
with $\mathcal{H}$ being the Hamiltonian in absence of impurity
potentials.  

In the correlation function, Eq.(\ref{correl}), the
impurity averaging products $\left\langle
  \mathcal{G}^\text{R}\mathcal{G}^\text{R}\right\rangle$ and
$\left\langle \mathcal{G}^\text{A}\mathcal{G}^\text{A}\right\rangle$ are neglected 
as they, in comparison with $\left\langle \mathcal{G}^\text{R}\mathcal{G}^\text{A}\right\rangle$, do not exhibit poles in the complex plane and are smaller by a factor $\left (\epsilon_F \tau\right )^{-1}$.\cite{Rammer2004,Bruus2002}  
The sum in Eq.~(\ref{correl})  averages over all intermediate electron momenta $\mathbf{q}$ of the scattering events. 
As stated before, we assume diffusive motion not only along the cylinder axis, but also along the circumference. 
This assumption holds true as long as the electrons' mean-free path is much smaller than the circumference of the nanowire.
Such situation is similar to a disordered planar 2D system and therefore we will treat it analogously.\cite{wenkdiss} 
As a consequence, the electron momentum $\mathbf{q}$ is considered as a continuous variable and replaced by the Fermi
velocity $\mathbf{q}=m \mathbf{v}_F$.
With this, 
we average over all directions of the in-plane momentum and the azimuthal angles.  
It is worth to mention, that in flat quantum wires, the specular scattering at the lateral confinement requires conservation of the spin current which yields an additional boundary condition for the Cooperon equation.\cite{Wenk2010,meyer}
This, however, does not apply to periodic boundary conditions and is therefore irrelevant for the tubular system.\cite{wenkdiss}

Defining the in-plane velocity $\mathbf{v}_\parallel
=(v_\phi,v_z)^\top=
v_\parallel\left(\cos(\vartheta),\sin(\vartheta)\right)^\top$ of the
cylinder's tangent space
the Eq.~(\ref{cooperon1}) simplifies  to
\begin{equation}
\hat{\mathcal{C}}(\mathbf{\hat{Q}})=\,\tau\left(1-\int_0^{2\pi} \frac{d\phi}{2\pi}\int_0^{2\pi} \frac{d\vartheta}{2\pi}\frac{1}{1-i \tau \hat{\Sigma}}\right)^{-1}
\label{cooperon2}
\end{equation}
where
\begin{equation}
\hat{\Sigma}=\,\mathcal{H}(\mathbf{\hat{Q}}-m \mathbf{v}_F,\boldsymbol{\sigma})-\mathcal{H}(m \mathbf{v}_F,\boldsymbol{\sigma'}).
\label{Sigma}
\end{equation}
The integral represents the averaging over the azimuth angle $\phi$ of the cylinder and the angle $\vartheta$ between the in-plane velocity components $v_\phi$ and $v_z$ in the tangent space corresponding to a certain angle $\phi$.
A more detailed derivation is given in App.~\ref{app:corr}.
In the following, we assume the ratio $\kappa\equiv\braket{v_r}/(i v_\parallel)$ to be small.
This holds true for $8\pi n_{2\text{D}}R^2\gg 1$
which can be seen when rewritten in terms of the 2D electron density $n_{2D}$ as $\kappa^2=(4m^2R^2v_F^2+1)^{-1}=(8\pi n_\text{2D}R^2+1)^{-1}$ by means of the relation $v_\text{F}=\sqrt{2\pi n_\text{2D}}/m$.
Therefore, we can approximate $v_F^2=\braket{v_r}^2+v_\parallel^2\approx v_\parallel^2$ and $D_e\approx\tau v_\parallel^2/2$.

A very important experimental tool to extract SOC strength are
magnetoconductivity measurements.\cite{Wirthmann2006, Schapers2006,
  PhysRevB.73.241311, Liang2009}  These measurements detect
the conductivity as a function of small perpendicular
magnetic fields which break the time-reversal symmetry as the
electron's wave function gains an Aharonov-Bohm phase and thereby
destroy the phase coherence.
Former approaches in
2D~\cite{nagaoka,Iordanskii1994,Pikus1995,Knap1996} dealt with
magnetic fields non-perturbatively in the basis of Landau bands.
However, since we are only interested in the behavior at small
magnetic fields, the Landau basis is not an appropriate choice.
We include small magnetic fields
$\mathbf{B}=\nabla\times\mathbf{A}$ purely by the principle of minimal
coupling and substitute the momenta $\mathbf{Q}\rightarrow\mathbf{Q}+2e
\mathbf{A}$.  We choose the magnetic field $\mathbf{B}=
B\,\mathbf{\hat{y}}$ which is related to the vector potential $\mathbf{A}$, here, represented
in Landau gauge as
\begin{align}
\mathbf{A}={}&-B x \mathbf{\hat{z}}=-B R \cos(\phi) \mathbf{\hat{z}}.
\label{vectorpot}
\end{align}
On the cylinder surface, the magnetic field as well as the vector
potential are inhomogeneous and the vector potential has no
out-of-plane component.

If we drop all terms in $\hat{\Sigma}$ that do not contain the Fermi
velocity, which gives the dominant contribution, we obtain
\begin{align}
\hat{\Sigma}\approx-\left(\mathbf{\hat{Q}}+2e\mathbf{A}+2 m \boldsymbol{\hat{a}}\boldsymbol{S}\right)\mathbf{v}_F.
\label{Sigma2}
\end{align}
The matrix $\boldsymbol{\hat{a}}$ contains the Rashba and Dresselhaus SOC and is listed in App.~\ref{app:socmatrices} for the different growth directions.
Furthermore, we define the total electron spin vector $\boldsymbol{S}$
which is
\begin{equation}
\boldsymbol{S}=\frac{1}{2}(\boldsymbol{\sigma}\otimes\mathbbm{1}+\mathbbm{1}\otimes\boldsymbol{\sigma'})
\label{S}
\end{equation}  
in the basis $\ket{s_1,s_2}$.
A more suitable choice for the spin matrix representation is the
singlet-triplet basis $\ket{s,m_s}$, though. 
Here, the total spin quantum number is labeled by $s\in\{0,1\}$, where $s=0$ defines the singlet $(\mathcal{S})$ and $s=1$ the triplet $(\mathcal{T})$ state, and the corresponding magnetic quantum numbers by $m_s$, where $m_s=0$ accounts for the singlet and $m_s\in\{\pm1,0\}$ for the triplet state.
The explicit form is given in App.~\ref{app:st}. 
Advantageously, in this representation the singlet and the triplet sectors decouple from each other and can be treated separately.

At last, we define the Cooperon Hamiltonian as
\begin{equation}
H_\mathcal{C}(\mathbf{\hat{Q}})\equiv\frac{1}{D_e\,\hat{\mathcal{C}}(\mathbf{\hat{Q}})}
\label{HCooperon}
\end{equation}
and perform the integral in Eq.~(\ref{cooperon2}) by Taylor expanding
the integrand to second order in
$\left(\mathbf{\hat{Q}}+2e\mathbf{A}+2 m
  \boldsymbol{\hat{a}}\boldsymbol{S}\right)$ and find the Cooperon
Hamiltonian in units of $Q_{so}^2=(2m\beta)^2$ as

\begin{widetext}
\begin{align}
\frac{H_\mathcal{C}^\xi(\boldsymbol{\hat{\mathcal{Q}}})}{Q_{so}^2}={}&\left(\mathcal{Q}_\phi^2+\mathcal{Q}_z^2+\mathcal{B}^2R_{so}^2\right)\,\mathbbm{1}_{4\times4}
-2\left(\lambda_1+\frac{\lambda_2}{\pi}\right)\mathcal{Q}_\phi S_z-\frac{1}{2}\lambda_2\mathcal{Q}_z S_x
+\lambda_1\left[\frac{2}{3\pi}\lambda_2\left(2S_x^2+S_y^2+3S_z^2\right)-\sqrt{2}\mathcal{B}R_{so}S_y\right]\notag\\
{}&+\lambda_1^2\left[\frac{1}{2}\left(S_x^2+S_y^2\right)+S_z^2\right]
+\frac{\lambda_2^2}{16}\left[3S_x^2+S_y^2+4\left(1-2\kappa^2\right)S_z^2\right]-\frac{2\sqrt{2}}{3\pi}\lambda_2\mathcal{B}R_{so}S_y+\mathcal{F}^\xi.
\label{HC}
\end{align}
The terms $\mathcal{F}^\xi$ result from Dresselhaus SOC and thus depend on the growth direction $\xi$ of the nanowire as
\begin{align}
\mathcal{F}^\text{[001]}&=\, \frac{1}{16}\left[ \left(5-82 \kappa^2\right)\left(S_x^2+S_y^2\right)+8S_z^2\right]+\frac{\lambda_2}{8}\left(3-2\kappa^2\right)\left\{S_x,S_z\right\},\label{HC001}\\
\mathcal{F}^\text{[111]}&=\,\frac{1}{8}\left(S_x^2+S_y^2+6S_z^2\right)-\frac{1}{12}\left(S_x^2+S_y^2+163S_z^2\right)\kappa^2+\frac{\lambda_2}{4\sqrt{3}}
\left[\left(1-2\kappa^2\right)\left\{S_y,S_z\right\}-\sqrt{2}\left\{S_x,S_y\right\}\right],\label{HC111}\\
\mathcal{F}^\text{[110]}&=\,\frac{3}{2} \mathcal{Q}_z S_x +\frac{1}{64}\left[38 S_x^2 + 32 S_y^2 + \left(41 - 730 \kappa^2\right)S_z^2\right] +\frac{\lambda_2}{16}\left[\left(2\kappa^2-1\right)S_z^2-7 S_x^2-4 S_y^2\right]\label{HC110}.
\end{align}
Here, we define the dimensionless parameters
\begin{equation}
\mathcal{Q}_i=\,\frac{Q_i}{Q_{so}},\quad
\lambda_1=\,\frac{\alpha_\text{int}}{\beta},\quad
\lambda_2=\,\frac{\alpha_\text{ext}}{\beta},\quad
\kappa=\,\frac{\braket{v_r}}{i v_\parallel},\quad
\mathcal{B}=\,\frac{\sqrt{2} e B}{Q_{so}^2},\quad
R_{so}=\,R\, Q_{so}.
\label{CooperonDefs}
\end{equation}
\end{widetext}
Note that in Eq.~(\ref{HC}) we neglect the terms
\begin{equation}
\frac{1}{D_e Q_{so}^2}\left(i \braket{Q_r} \braket{v_r}+\braket{Q_r}^2 \braket{v_r}^2 \tau\right)\mathbbm{1}_{4\times4}.
\label{HCapprox}
\end{equation}
This is justified as the first term is proportional to $\tau^{-1}$
since $D_e\propto\tau$, which is in accordance with the case without
SOC as shown in Ref.~\onlinecite{Bruus2002}.
The second term can be dropped because
$\braket{Q_r}^2 \braket{v_r}^2\propto R^{-4}$ and thus it is very small for a large radius.  
The Cooperon Hamiltonian is therefore Hermitian and we will discuss its spectrum hereafter.

\subsection{Spectrum Analysis}

\subsubsection{Analytical Expressions for the Eigenvalues}

In general, there is no simple analytical expression for the eigenvalues of the full Cooperon Hamiltonian in  Eq.~(\ref{HC}). 
Partly, this is attributed to the reduction of symmetry by the external Rasbha, the Dresselhaus SOC for wires along $[110]$ and the magnetic field.
Yet, we can provide solutions of simple structure for certain particular situations.
This will be useful for estimating the spin relaxation rates and determining the conductivity correction in Sec.~\ref{subsec:corr}.

Since the magnetic field is considered to be small, it is reasonable to neglect the
off-diagonal terms $\propto \mathcal{B}$ in Eq.~(\ref{HC}). 
The magnetic field will, hence, merely cause a shift $\mathcal{B}^2R_{so}^2$ of the entire spectrum.
Note that this is equivalent to
treating the magnetic field by means of a magnetic phase shift rate
$\tau_B$ that breaks the time-reversal invariance, which gives for
diffusive wire cross-sections $1/\tau_B=\,2 D_e e^2 R^2 B^2$. \cite{km02_3,MazzarelloDiss} 

If the surface conductive channel is a consequence of Fermi level pinning, the internal Rashba and Dresselhaus SOC can be comparably large and compete with each other. 
Yet, here we consider the Dresselhaus SOC to be the dominant mechanism as it strongly depends on the confinement due to the matrix element $\braket{k_r^2}$ which is only a few tens of nanometers in a realistic nanowire\cite{Bringer2011,BloemersDiss,Hernandez2010}.
Also, in a situation where the gate is wrapped around the nanowire the resulting field is collinear to the internal field and therefore renormalizes the internal Rashba coefficient.\cite{Storm2012,Liang2012,Karlstroem2008} 
Moreover, in core/shell systems the band bending can be much lower.\cite{Bloemers2013}
Thus, the Rashba SOC will constitute a small perturbation whereas the internal outweighs the external as the latter is due to a gate voltage and can be chosen arbitrarily small. 
 
In line with this, considering the high-symmetry growth directions $[001]$ and $[111]$ we can provide an approximate solution for the band structure by neglecting all off-diagonal elements proportional to $\lambda_2$.
We stress that these eigenvalues are \textit{exact} for vanishing external
Rashba contribution, i.e., $\lambda_2=0$ (and neglected off-diagonal magnetic field terms).
In case of the low-symmetry direction $[110]$ owing to finite off-diagonal Dresselhaus terms we find an exact solution only by neglecting \textit{all} Rashba contributions, i.e., $\lambda_1=\lambda_2=0$.
Thus, the spectrum of the Cooperon Hamiltonian is given by
\begin{align}
E_\mathcal{S}/Q_{so}^2&={}\mathcal{Q}_\phi^2+\mathcal{Q}_z^2+\mathcal{B}^2R_{so}^2,\label{Esinglet}\\
E^\xi_\chi/Q_{so}^2&\approx{}E_\mathcal{S}/Q_{so}^2+\mathcal{M}^\xi_\chi,\label{Etriplet}
\end{align}
where for the high-symmetry directions we obtain
\begin{align}
\mathcal{M}^{[001]}_{\mathcal{T}_0}={}&f_0+\frac{5}{8}-\frac{41}{4}\kappa^2,\label{001M0}\\
\mathcal{M}^{[001]}_{\mathcal{T}_{\pm1}}={}&f_\pm+\frac{13}{16}-\frac{41}{8}\kappa^2\label{001M+-},
\end{align}
and
\begin{align}
\mathcal{M}^{[111]}_{\mathcal{T}_0}={}&f_0+\frac{1}{4}-\frac{1}{6}\kappa^2,\label{111M0}\\
\mathcal{M}^{[111]}_{\mathcal{T}_{\pm1}}={}&f_\pm+\frac{7}{8}-\frac{163}{12}\kappa^2\label{111M+-},
\end{align}
with
\begin{align}
f_0={}&\,\lambda_1^2+\frac{2}{\pi}\lambda_1\lambda_2+\frac{1}{4}\lambda_2^2,\\
f_\pm={}&\frac{3}{2}\lambda_1^2+\frac{1}{2}\left(\frac{3}{4}-\kappa^2\right)\lambda_2^2+\frac{3}{\pi}\lambda_1\lambda_2\notag\\
{}&\pm2\left(\lambda_1+\frac{1}{\pi}\lambda_2\right)\mathcal{Q}_\phi
\label{eq:EVtot}
\end{align}
and for the low-symmetry direction without Rashba SOC
\begin{align}
\mathcal{M}^{[110]}_{\mathcal{T}_0}={}&\frac{73}{64}-\frac{10}{64}\kappa^2,\label{110M0}\\
\mathcal{M}^{[110]}_{\mathcal{T}_{\pm1}}={}&\frac{1}{128}\Bigg[149-730\kappa^2\label{110M+-}\notag\\
{}&\pm\sqrt{36864 \mathcal{Q}_z^2+\left(9-730\kappa^2\right)^2}\Bigg].
\end{align}
Here, the Cooperon momentum operator $\boldsymbol{\hat{\mathcal{Q}}}$ is expressed in the basis given in Eqs.~(\ref{eigenfcthollowH0phi}) and (\ref{eigenfcthollowH0z}).
Thus, the Cooperon momentum along the cylinder axis $\mathcal{Q}_z$ is quasi-continuous and the transverse Cooperon momentum becomes $\mathcal{Q}_\phi=n/R_{so}$, where $n$ is the number of the transverse Cooperon mode. 

\begin{figure}[t]
\includegraphics[width=.95\columnwidth]{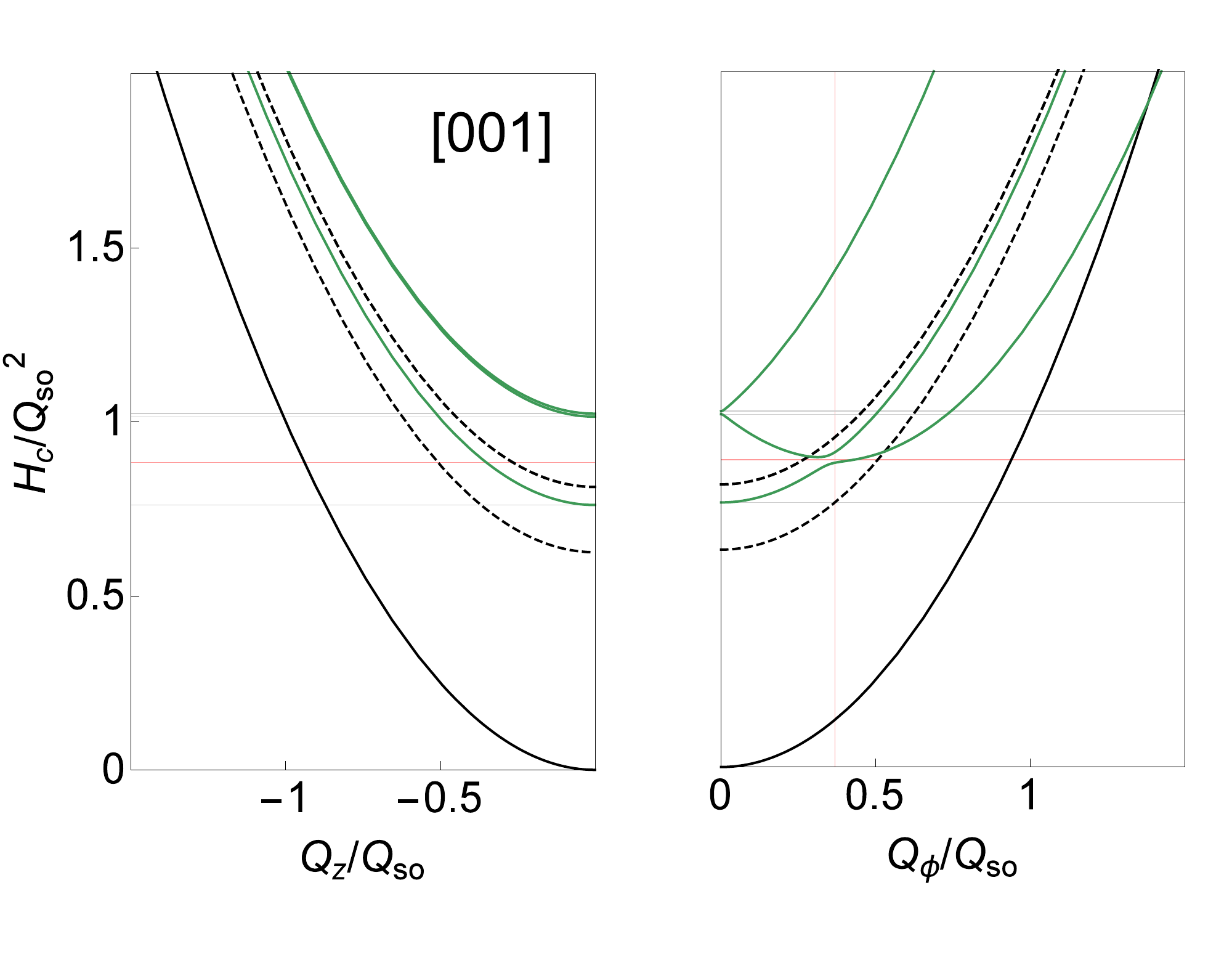}
\caption{(Color online) Spectrum of the Cooperon Hamiltonian for $[001]$ nanowires with parameter configurations $\lambda_1=0.4$ and $\lambda_2=-0.1$ for $\mathcal{B}=\kappa=0$ (green). Dashed lines correspond to vanishing Rashba SOC and black solid line to the singlet mode. The grid lines are plotted by use of the approximate formulas $\Delta_\chi^{[001]}$ (gray) and $\delta_\phi^{[001]}$ for $\mathcal{Q}_\phi^\text{min,[001]}$ (red). The grid lines are plotted using the approximate formulas derived in Sec.~\ref{subsubsec:SRG}.} 
\label{CooperonSpec001}
\end{figure}

\begin{figure}[t]
\includegraphics[width=.95\columnwidth]{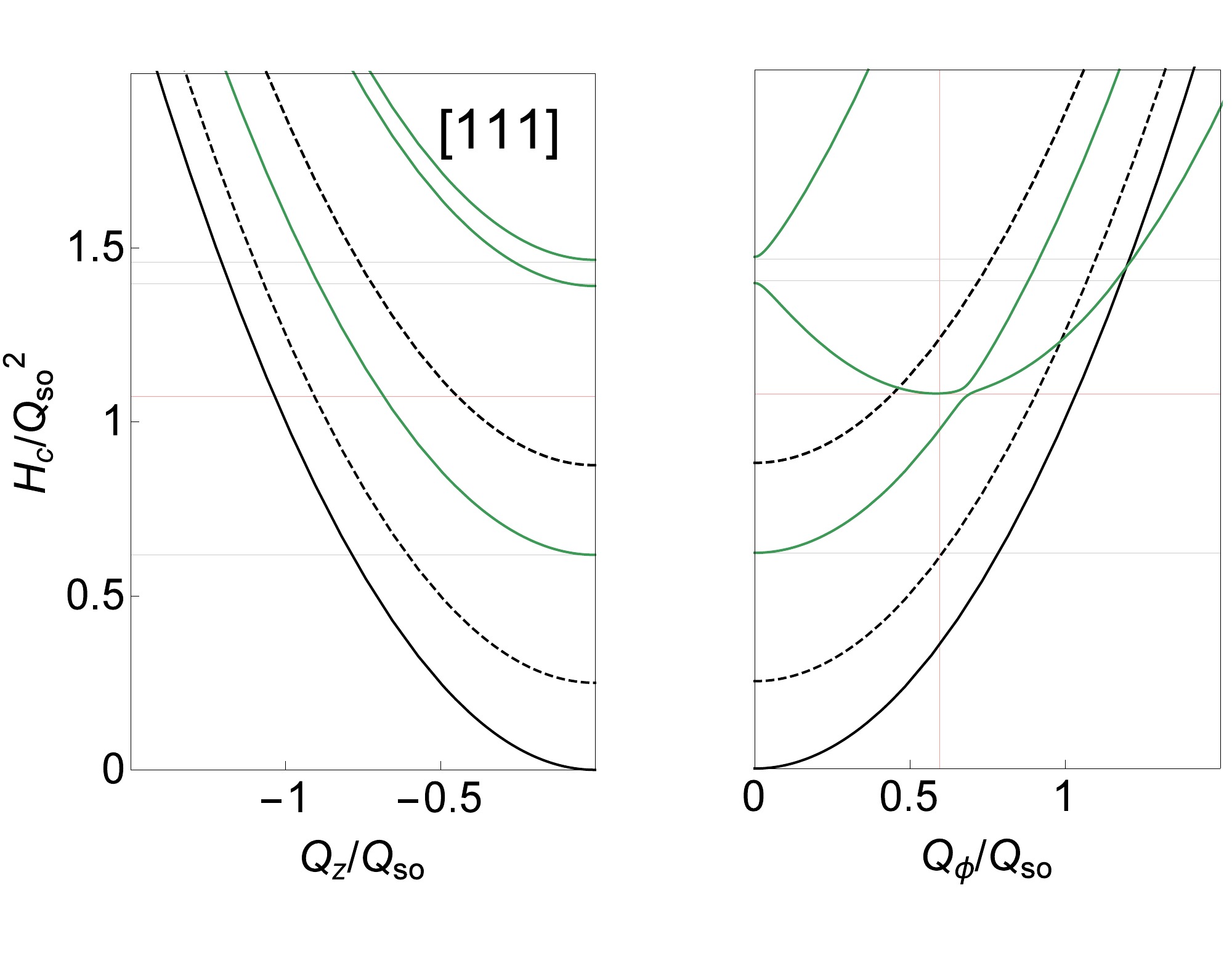}
\caption{(Color online) Spectrum of the Cooperon Hamiltonian for $[111]$ nanowires with parameter configurations $\lambda_1=0.5$ and $\lambda_2=0.3$ for $\mathcal{B}=\kappa=0$ (green). Dashed lines correspond to vanishing Rashba SOC and black solid line to the singlet mode. The grid lines are plotted by use of the approximate formulas $\Delta_\chi^{[111]}$ (gray) and $\delta_\phi^{[111]}$ for $\mathcal{Q}_\phi^\text{min,[111]}$ (red). The grid lines are plotted using the approximate formulas derived in Sec.~\ref{subsubsec:SRG}.}
\label{CooperonSpec111}
\end{figure}

\begin{figure}[t]
\includegraphics[width=.95\columnwidth]{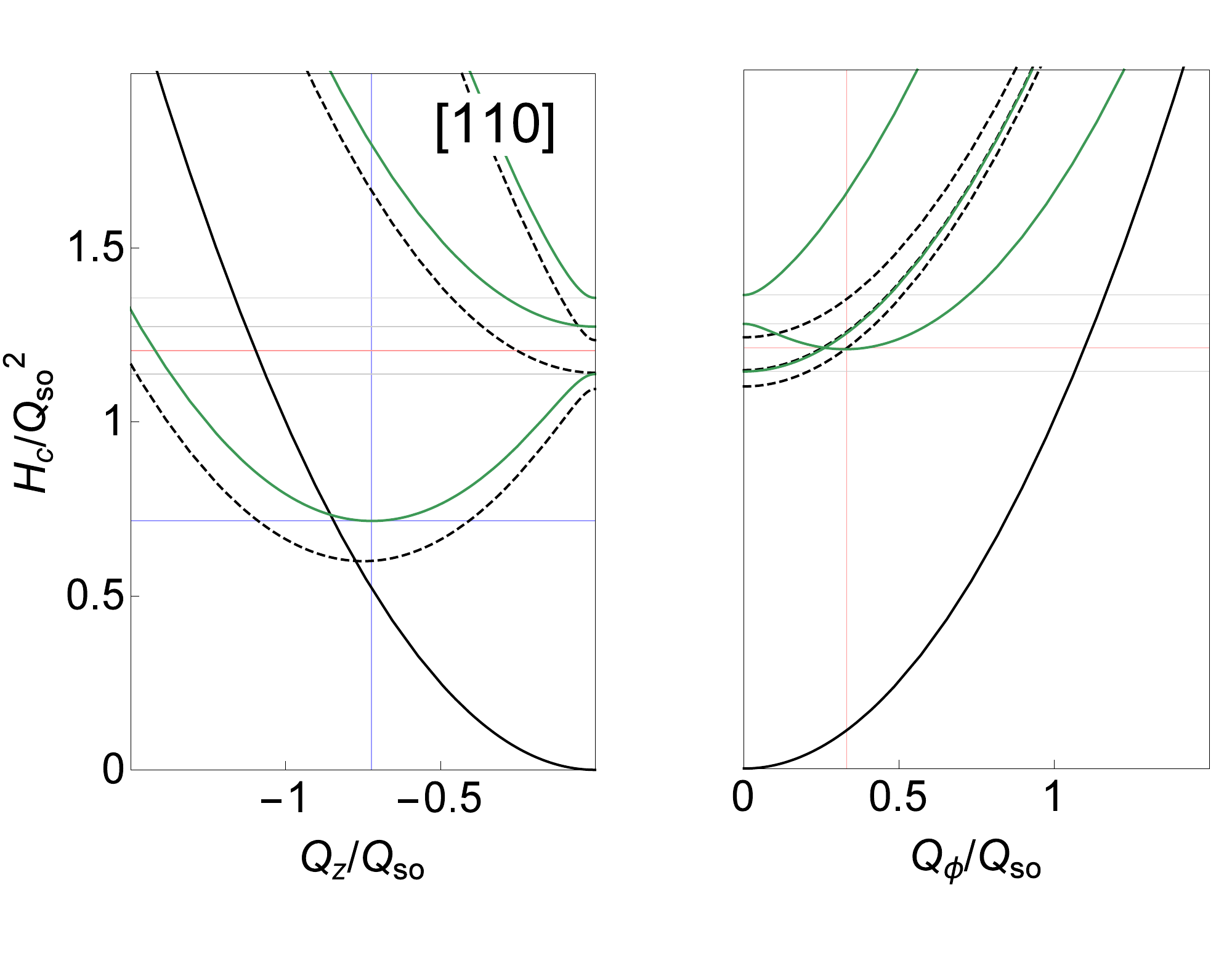}
\caption{(Color online) Spectrum of the Cooperon Hamiltonian for $[110]$ nanowires with parameter configurations $\lambda_1=0.3$ and $\lambda_2=0.1$ for $\mathcal{B}=\kappa=0$ (green). Dashed lines correspond to vanishing Rashba SOC and black solid line to the singlet mode. The grid lines are plotted by use of the approximate formulas $\Delta_\chi^{[110]}$ (grey), $\delta_\phi^{[110]}$ for $\mathcal{Q}_\phi^\text{min,[110]}$ (red) and $\delta_z^{[110]}$ for $\mathcal{Q}_z^\text{min,[110]}$ (blue). The grid lines are plotted using the approximate formulas derived in Sec.~\ref{subsubsec:SRG}.}
\label{CooperonSpec110}
\end{figure}

\begin{figure}[t]
\includegraphics[width=.95\columnwidth]{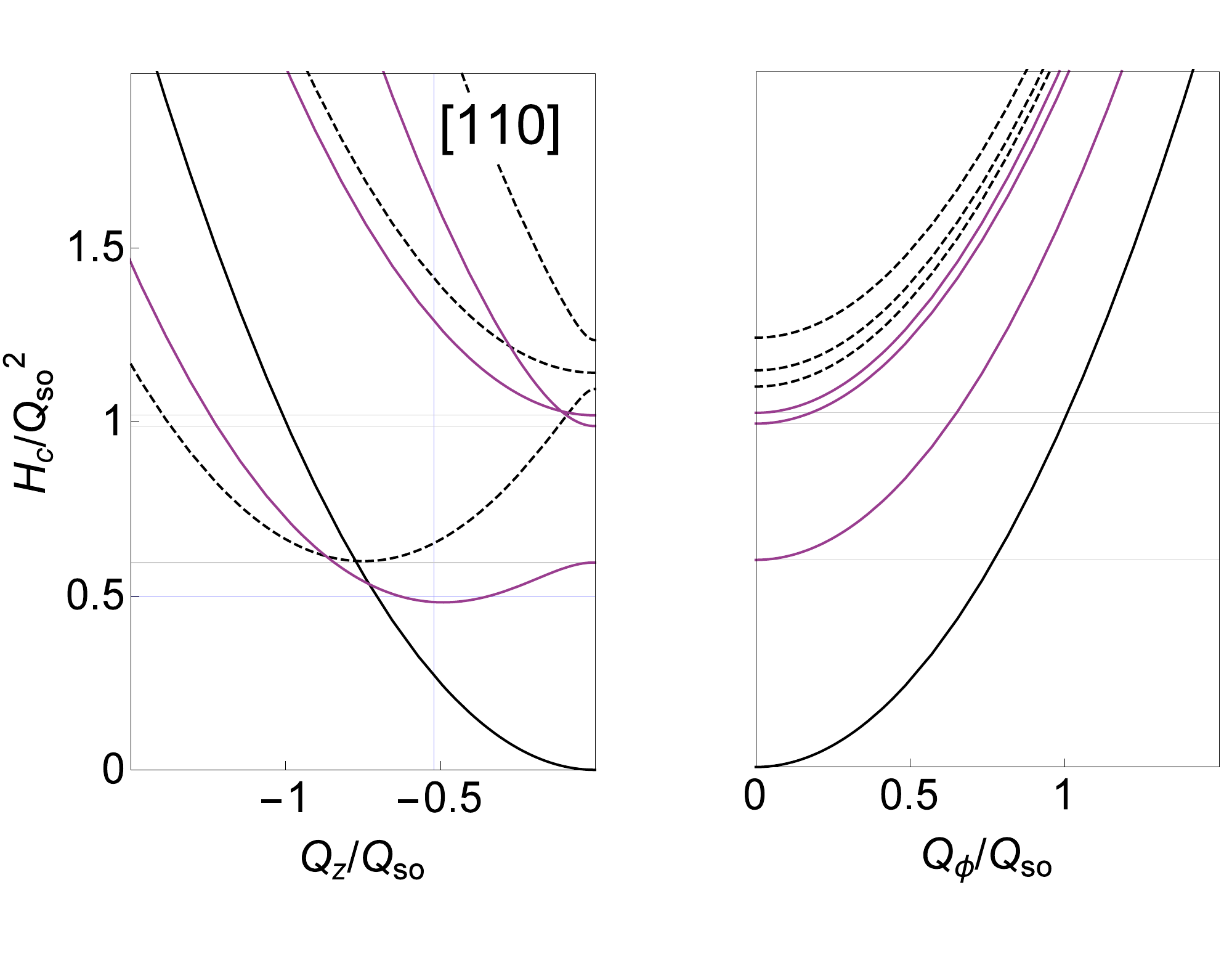}
\caption{(Color online) Spectrum of the Cooperon Hamiltonian for $[110]$ nanowires with optimal parameter configurations for a lowest possible gap along $\mathcal{Q}_z$ with $\lambda_1=-0.901$ and $\lambda_2=0.305$ for $\mathcal{B}=\kappa=0$ (green). Dashed lines correspond to vanishing Rashba SOC and black solid line to the singlet mode. The grid lines are plotted by use of the approximate formulas $\Delta_\chi^{[110]}$ (grey) and $\delta_z^{[110]}$ for $\mathcal{Q}_z^\text{min,[110]}$ (blue). The grid lines are plotted using the approximate formulas derived in Sec.~\ref{subsubsec:SRG}.}
\label{CooperonSpec110opt}
\end{figure}

The exact energy spectra of the Cooperon Hamiltonian Eq.~(\ref{HC}) are displayed in
Figs.~\ref{CooperonSpec001}, \ref{CooperonSpec111}, \ref{CooperonSpec110} and \ref{CooperonSpec110opt} for
$\mathcal{B}=\kappa=0$. 
For better perceptibility we illustrate $\mathcal{Q}_\phi$ as a continuous quantity, which corresponds to the case where $R_{so}\gg 1$. 
In all figures the black solid line corresponds to the singlet mode,
which is independent of the SOC, and the black dashed lines to the case where Rashba SOC is absent.

\subsubsection{Spin Relaxation Gaps}\label{subsubsec:SRG}

As the spectrum of the Cooperon and the spin diffusion equation are identical as far as the time-reversal symmetry is not broken, i.e., $B=0$, the minima of the triplet eigenvalues of the Cooperon Hamiltonian are direct measures of the spin relaxation rate and thus of particular interest.\cite{Wenk2010,Schwab2006}
In Eq.~(\ref{HC}) all terms that are linear in momentum $\mathcal{Q}$ can shift the minimum of two triplet eigenmodes to a finite momentum.
Any mode that is gapless at finite $\mathcal{Q}$ reveals a persistent spin helix which has been demonstrated in 2DEGs for a certain ratio of linear Dresselhaus and Rashba SOC strength.\cite{wenkbook}
However, only in growth direction $[110]$ there is a $\mathcal{Q}$-dependent Dresselhaus term.
This reflects an earlier statement that the interplay with Rashba cannot be used to control the minimum and suppress the spin relaxation for $\braket{001}$ and $\braket{111}$ nanowires.
Also, we see from Eqs.~(\ref{HC001})-(\ref{HC110}) that even for $\mathcal{Q}$-independent terms  there is no coupling between Dresselhaus and internal Rashba SOC in any growth direction.
We only find a coupling between Dresselhaus and external Rashba as well as internal and external Rashba.
These $\mathcal{Q}$-independent terms cause a positive shift of the triplet spectrum and thereby an insuppressible spin relaxation.
Hence, a gapless mode cannot be found.
This contradicts the conjecture of previous authors, Refs.~\onlinecite{Rainis2014,Dhara2009,Weperen2015,Hansen2005,Nadj2012,
Roulleau2010}, that Dresselhaus SOC in $[111]$ is absent and, hence, cannot cause spin relaxation.
In the following, we analyze the position and value of the minima for $\mathcal{B}=0$ which can be related to spin relaxation rates. 
The latter we will denote as spin relaxation gaps.

First of all, we note that $\kappa$ is the only parameter which lowers the
triplet eigenenergies at $\boldsymbol{\mathcal{Q}}=0$.  
It is remarkable as this quantity depends on the radius $R$ of the nanowire.
Nevertheless, it is assumed to be small since we consider $n_{2\text{D}}R^2\gg 1$.
Therefore, we will neglect $\kappa$ for simplicity in the following discussion.

Since the Rashba SOC constitutes a small perturbation, we can estimate the spin relaxation gaps $\Delta^\xi_\chi\equiv E^\xi_\chi\left(\boldsymbol{\mathcal{Q}}=0\right)/Q_{so}^2$ by Taylor expanding the exact eigenvalues in terms of $\lambda_i$  as
\begin{align}
\Delta^{[001]}_{\mathcal{T}_{0}}\approx{}&\lambda_1^2+\frac{2}{\pi}\lambda_1\lambda_2+\frac{1}{16}\lambda_2^2+\frac{5}{8},\label{Delta001}\\
\Delta^{[001]}_{\mathcal{T}_{\pm1}}\approx{}&\frac{3}{2}\lambda_1^2+\frac{9\mp1}{3\pi}\lambda_1\lambda_2+\frac{15\pm1}{32}\lambda_2^2+\frac{13}{16},\\[10pt] 
\Delta^{[111]}_{\mathcal{T}_0}\approx{}&\lambda_1^2+\frac{2}{\pi}\lambda_1\lambda_2+\frac{29}{120}\lambda_2^2+\frac{1}{4},\label{Delta111}\\
\Delta^{[111]}_{\mathcal{T}_{\pm1}}\approx{}&\frac{3}{2}\lambda_1^2+\frac{3}{\pi}\lambda_1\lambda_2+\frac{91}{240}\lambda_2^2+\frac{7}{8}\pm\frac{\left|\lambda_2\right|}{4\sqrt{6}},\\[10pt] 
\Delta^{[110]}_{\mathcal{T}_0}\approx{}&\lambda_1^2+\frac{2}{\pi}\lambda_1\lambda_2+\frac{1}{4}\lambda_2^2-\frac{11}{16}\lambda_2+\frac{35}{32},\\
\Delta^{[110]}_{\mathcal{T}_{\pm1}}\approx{}&\frac{3}{2}\lambda_1^2+\frac{9\pm1}{3\pi}\lambda_1\lambda_2+\frac{6\pm1}{16}\lambda_2^2\notag\\
{}&-\frac{13\pm3}{32}\lambda_2+\frac{76\pm3}{64}.
\end{align}
The twofold degeneracy of the eigenvalues $E^\xi_{\mathcal{T}_{\pm1}}$ at $\boldsymbol{\mathcal{Q}}=0$ is lifted   for $[110]$ nanowires and also, independent of the growth direction,  in presence of an external Rashba contribution owing to the lower symmetry.
An important observation at this point is that in absence of Rashba SOC the lowest spin relaxation gap is given for $[111]$ nanowires by $\Delta^{[111]}_{\mathcal{T}_0}=1/4$.
Thus, it is reasonable to assume that the spin relaxation due to Dresselhaus SOC is lowest for nanowires grown along $[111]$.

Analogously, let us call the spin relaxation gap, where a minimum in one of the triplet modes occurs at finite values of $\mathcal{Q}_i$, $\delta^\xi_i\equiv E^\xi_{\mathcal{T}_\text{min}}\left(\mathcal{Q}_i=\mathcal{Q}_i^{\text{min},\xi}\right)/Q_{so}^2$.
For arbitrary $\xi$ we can approximately locate the position of the minima at finite $\mathcal{Q}_\phi$ at $\mathcal{Q}_\phi^{\text{min},\xi}=\pm\left(\lambda_1+\lambda_2/\pi\right)$ by neglecting all off-diagonal elements.\footnote{For $[110]$ this assumption is rather crude due to off-diagonal Dresselhaus contributions.}
The spin relaxation gap $\delta^\xi_\phi$ at this position and in this approximation for $\mathcal{Q}_z=0$ is about
\begin{align}
\delta^{[001]}_\phi\approx{}&a+\frac{13}{16},\\
\delta^{[111]}_\phi\approx{}&a+\frac{7}{8},\\
\delta^{[110]}_\phi\approx{}&a-\frac{13}{32}\lambda_2+\frac{19}{16},\\
\intertext{where}\notag\\
a={}&\frac{1}{2}\lambda_1^2+\frac{1}{\pi}\lambda_1\lambda_2+\left(\frac{3}{8}-\frac{1}{\pi^2}\right)\lambda_2^2.
\end{align}
Depending on Rashba SOC, we can obtain a situation where the lowest minima are at finite values of $\mathcal{Q}_\phi$.
To linear order in $\lambda_2$ the domain $\mathcal{P}^\xi$ where the lowest minimum is at $\boldsymbol{\mathcal{Q}}=0$, that is, $\lambda_1\in\mathcal{P}^\xi$, is 
\begin{align} 
\mathcal{P}^{[001]}={}&\left(-\sqrt{\frac{3}{8}}-\frac{\lambda_2}{\pi},\sqrt{\frac{3}{8}}-\frac{\lambda_2}{\pi}\right),\\ \mathcal{P}^{[111]}={}&\left(-\frac{\sqrt{5}}{2}-\frac{\lambda_2}{\pi},\frac{\sqrt{5}}{2}-\frac{\lambda_2}{\pi}\right),\\ \mathcal{P}^{[110]}={}&\left(-\sqrt{\frac{3}{4}}-\left[\frac{1}{\pi}+\frac{3\sqrt{3}}{8}\right]\lambda_2,\right.\notag\\
{}&\left.\sqrt{\frac{3}{4}}-\left[\frac{1}{\pi}-\frac{3\sqrt{3}}{8}\right]\lambda_2\right).
\end{align}
However, in any case where the lowest minimum is at finite $\mathcal{Q}_\phi$, this spin relaxation gap is larger than compared to the case without Rashba SOC.
Therefore, it is reasonable to say that the spin direction of the long-lived spin states is homogeneous in coordinate space.


Contrary to the other cases, the $[110]$ direction reveals also a shifted minimum along $\mathcal{Q}_z$.
As a result, the states with the longest spin lifetime are of helical nature along the wire axis.\cite{wenkdiss}
By expanding the exact eigenvalues to first order in $\lambda_i$ and setting $\mathcal{Q}_\phi=0$ we obtain 
\begin{equation}
\mathcal{Q}_z^{\text{min},[110]}=\pm\frac{3\sqrt{255}}{64}\mp\frac{87}{64}\sqrt{\frac{3}{85}}\lambda_2,
\end{equation}
which yields the spin relaxation gap 
\begin{align}
\delta^{[110]}_z\approx{}&\frac{2455}{4096}-\frac{463}{2048}\lambda_2+\frac{79}{64}\lambda_1^2+\frac{21}{8\pi}\lambda_1\lambda_2+\frac{1093}{4096}\lambda_2^2.\label{delta110}
\end{align}
It is the lowest gap in the $[110]$ triplet spectrum until the Rashba contribution becomes very large, that is, $\left|\lambda_1\right|>1.45$ for pure internal or $\lambda_2<-28.4\vee\lambda_2>1.03$ for pure external Rashba SOC.
We find an optimal value of $\delta^{[110]}_z\approx0.498$ for $\lambda_1\approx-0.305$ and $\lambda_2\approx0.901$ within our approximations. 

In what follows, we apply the previously derived approximate formulas to compute the correction to the Drude conductivity.

\subsection{Correction to the Static Conductivity}\label{subsec:corr}
As shown in more detail in App.~\ref{app:st}, the sum over spin
indices in Eq.~(\ref{conductivity}) simplifies, in singlet-triplet
representation, to
\begin{align}
\Delta\sigma={}&\,\frac{e^2 }{\pi\mathcal{V}}\sum_{\mathbf{Q}}\left(\frac{1}{E^{\mathcal{S}}(\mathbf{Q})}-\sum_{i\in\{\pm1,0\}}\frac{1}{E^{\mathcal{T}}_i(\mathbf{Q})}\right),
\label{conductivity2}
\end{align}
where the $E_i^{j}$ are the eigenvalues of the Cooperon Hamiltonian, Eq.~(\ref{HC}).
Note the opposite sign of the singlet and triplet eigenvalues. The
dominance of the singlet or triplet sector determines whether the
conductivity correction results in WL or WAL.

Since the $\phi$ angular dependence is removed by the
integration over Fermi velocity $\mathbf{v}_F$ in Eq.~(\ref{cooperon2}), there is no coupling
between the transverse Cooperon modes.  
In experiment the infrared and ultraviolet divergence is eliminated due to a finite dephasing and elastic scattering time, $\tau_\phi$ and $\tau$, respectively.
Therefore, we insert a lower cutoff $c_\phi$ due to dephasing and an upper cutoff $c_\tau$ due to elastic scattering.

\begin{figure}[t]
\includegraphics[width=.97\columnwidth]{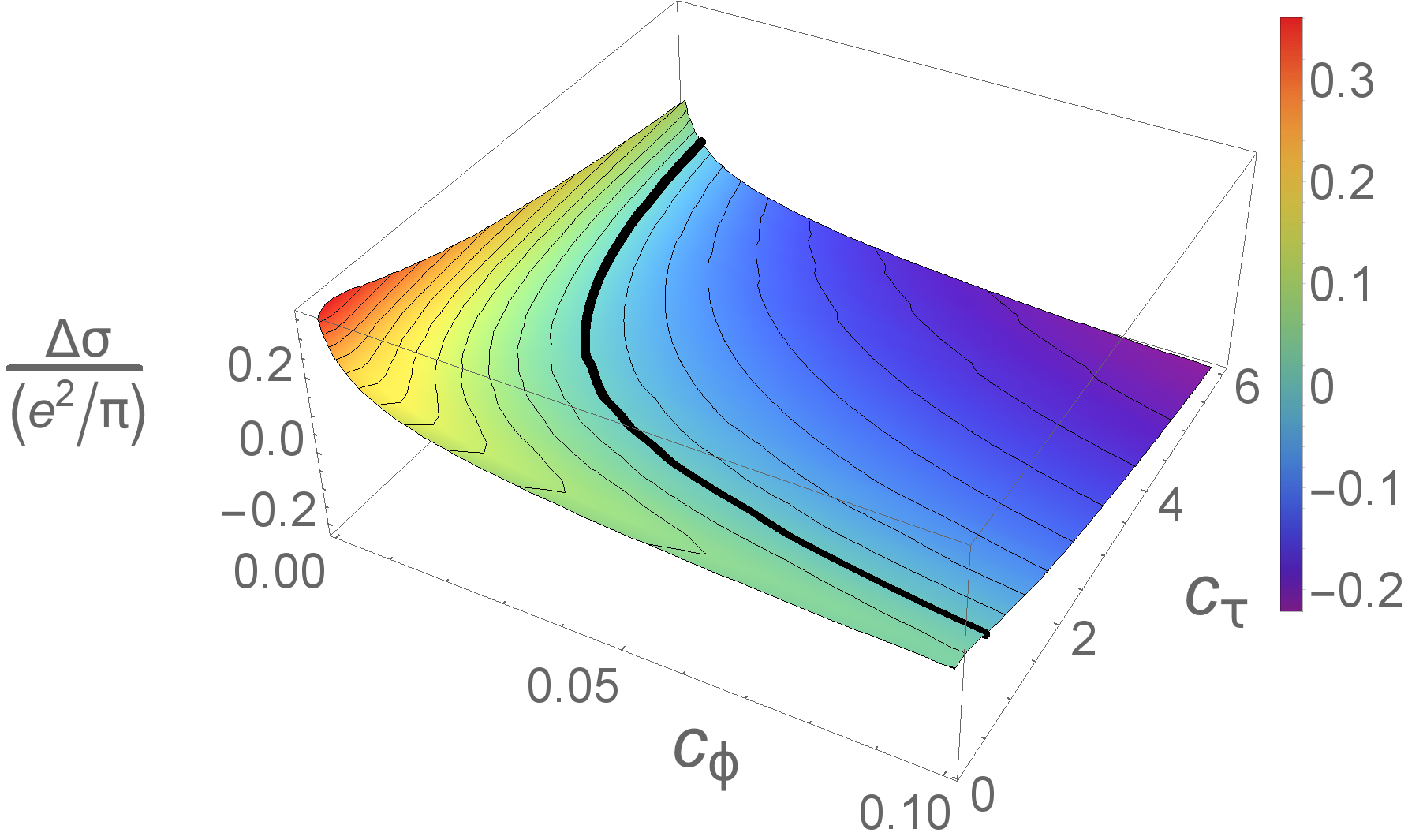}
\caption{(Color online) Crossover from WL to WAL in a $\braket{111}$ nanowire for $\mathcal{B}=\kappa=\lambda_2=0$, $\lambda_1=0.3$ and $R_{so}=30$.}
\label{pic:conductivity1}
\end{figure}

Consequently, in terms of $Q_{so}$ Eq.~(\ref{conductivity2}) becomes 
\begin{align}
\Delta\sigma={}&\,\frac{e^2 }{\pi}\frac{1}{2\pi^2 R_{so}}\sum_{n=-l_\text{max}}^{l_\text{max}}\times\notag\\
{}& \int_{0}^{\sqrt{c_\tau}}d\mathcal{Q}_z\left(\frac{1}{E^{\mathcal{S}}(\mathcal{Q}_z,n)/Q_{so}^2+c_\phi}\right.\notag\\
{}&\phantom{\int_{0}^{\sqrt{c_\tau}}d\mathcal{Q}_z}-\sum_{i\in\{\pm1,0\}}\frac{1}{E^{\mathcal{T}}_i(\mathcal{Q}_z,n)/Q_{so}^2+c_\phi}\bigg),
\label{conductivity3}
\end{align}
where
\begin{align}
l_\text{max}={}&\lfloor\sqrt{c_\tau}R_{so}\rfloor,\\
c_\tau={}&1/(D_e \tau Q_{so}^2),\\
c_\phi={}&1/(D_e \tau_\phi Q_{so}^2)
\end{align}
and $\lfloor...\rfloor$ denotes the next lower integer number.
For the growth directions $[001]$ and $[111]$, we can further simplify Eq.~(\ref{conductivity3}) if we consider the approximate eigenvalues of the Cooperon Hamiltonian. 
In this case, the integral can be computed analytically and yields
\begin{widetext}
\begin{align}
\Delta\sigma&={}\,\frac{e^2 }{\pi}\frac{1}{2\pi^2 R_{so}}\sum_{n=-l_\text{max}}^{l_\text{max}}\left\{\frac{\arctan\left(\sqrt{c_\tau}\left[E^\mathcal{S}(Q_z=0,n)/Q_{so}^2+c_\phi\right]^{-1/2} \right)}{\left[E^\mathcal{S}(Q_z=0,n)/Q_{so}^2+c_\phi\right]^{1/2}}\right.\notag\\
{}&\left.-\sum_{i\in\{\pm1,0\}}\frac{\arctan\left(\sqrt{c_\tau}\left[E^\mathcal{T}_i(Q_z=0,n)/Q_{so}^2+c_\phi\right]^{-1/2}\right)}{\left[E^\mathcal{T}_i(Q_z=0,n)/Q_{so}^2+c_\phi\right]^{1/2}}\right\},
\end{align}
\end{widetext}
with the approximate eigenmodes $E_i^j$ [Eqs.~(\ref{Esinglet}) and (\ref{Etriplet})] of the Cooperon Hamiltonian evaluated at $Q_z=0$.
%
%
%
\begin{figure}[t]
\includegraphics[width=.95\columnwidth]{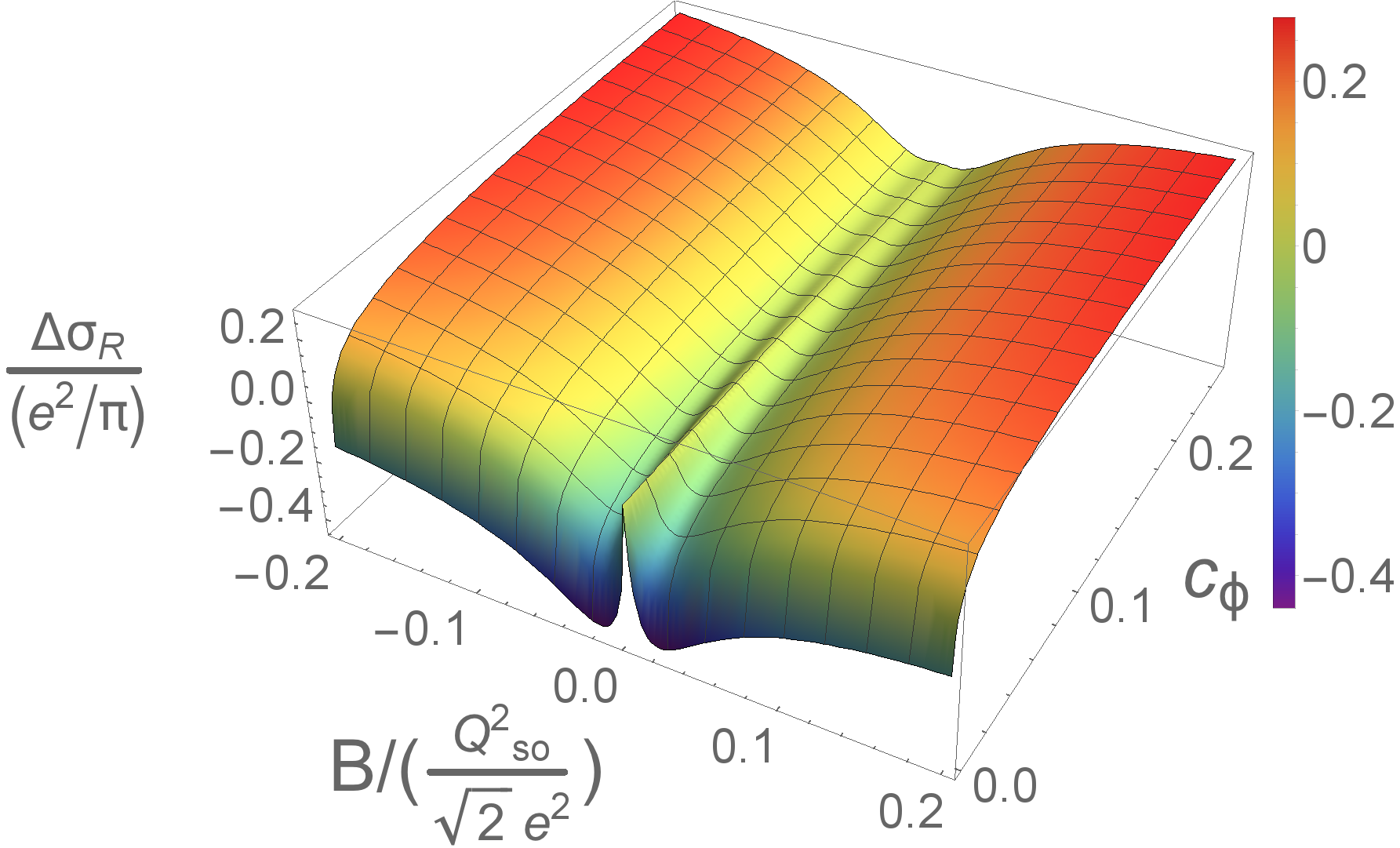}
\caption{(Color online) Relative magnetoconductivity $\Delta\sigma_R\equiv\Delta\sigma(B)-\Delta\sigma(B=0)$ in a $\braket{111}$ nanowire for $\kappa=\lambda_2=0$, $\lambda_1=0.3$, $R_{so}=30$ and $c_\tau=10$.}
\label{pic:conductivity3}
\end{figure}
In Fig.~\ref{pic:conductivity1}, we picture the conductivity correction
without magnetic and external electric field for a $\braket{111}$ nanowire. 
A crossover from WL to WAL appears depending on the dephasing time and the elastic scattering time.  The crossover from negative to positive magnetoconductivity is shown in
Figs. \ref{pic:conductivity3} which is due
to an increase of the lower cutoff $c_\phi$. 
Here, we defined the relative magnetoconductivity as $\Delta\sigma_R\equiv\Delta\sigma(B)-\Delta\sigma(B=0)$.
Note that this can also be achieved by reducing the SOC strength which is incapsulated in the
quantity $Q_{so}$.  It is worth mentioning that both crossovers
do not necessarily coincide.  Moreover, in contrast to a planar wire
with hard wall boundaries as shown in Ref.~\onlinecite{Wenk2010} we do
not find a crossover in dependency of the wire width $W$ which
corresponds to the circumference $2 \pi R$ of the tubular nanowire.
This is due to the fact that no motional narrowing occurs due to
periodic boundary conditions. \cite{wenkdiss}

By fitting theory to data from experiment one can extract the Rashba and Dresselhaus SOC strengths.
These are related to the spin relaxation rate $1/\tau_\text{s}$ (here, in units of eV) through 
\begin{align}
\frac{1}{\tau_s}={}&D_e E^\xi_\chi\Big|_{\mathcal{B}=0},
\label{spinrelaxtime}
\end{align}
where $E^\xi_\chi$ are the triplet eigenvalues of the Cooperon Hamiltonian without magnetic field.
The relaxation thus depends on the given spin state.
With the aid of the global minimum of the triplet spectrum one can estimate the minimal spin relaxation rate, though.
For the most systems it is reasonable to assume that those gaps are given by $\Delta^{[001]}_{\mathcal{T}_0}$, $\Delta^{[111]}_{\mathcal{T}_0}$, or $\delta^{[110]}_{z}$ in Eqs.~(\ref{Delta001}), (\ref{Delta111}) and (\ref{delta110}), depending on the growth direction of the nanowire.
For pure internal Rashba SOC, i.e., $\alpha_\text{ext}=\beta=0$, the spin relaxation rate at $Q=0$ yields 
\begin{align}
\frac{1}{\tau_s}={}&4m^2D_e\alpha_\text{int}^2,
\end{align}
which is identical to the case of a planar 2DEG as derived by D'yakonov \textit{et al.} in Ref.~\onlinecite{Dyakonov1986} and also found in early studies on WL/WAL\cite{Iordanskii1994,Knap1996}. 
We point out that for large Rashba SOC, the global minimum is shifted to finite momenta $Q\neq0$, though.
In case of pure internal Rashba SOC the resulting spin relaxation rate is about a factor $2$ smaller due to the gap $\delta_\phi\approx\lambda_1^2/2=\Delta_{\mathcal{T}_0}/2$, which was also noticed by Kettemann in Ref.~\onlinecite{Kettemann2007a}.
Owing to the discrete nature of $Q_\phi$ in our model, however, such states are not necessarily available in the given nanowire system.

In the last part of this section, we fit the derived formulas for the magnetoconductivity correction to experimental data of an exemplary semiconductor nanowire.

\subsection{Experimental Data Fitting: InAs Nanowire}\label{sec:fit}

\begin{figure}[t]
\includegraphics[width=.95\columnwidth]{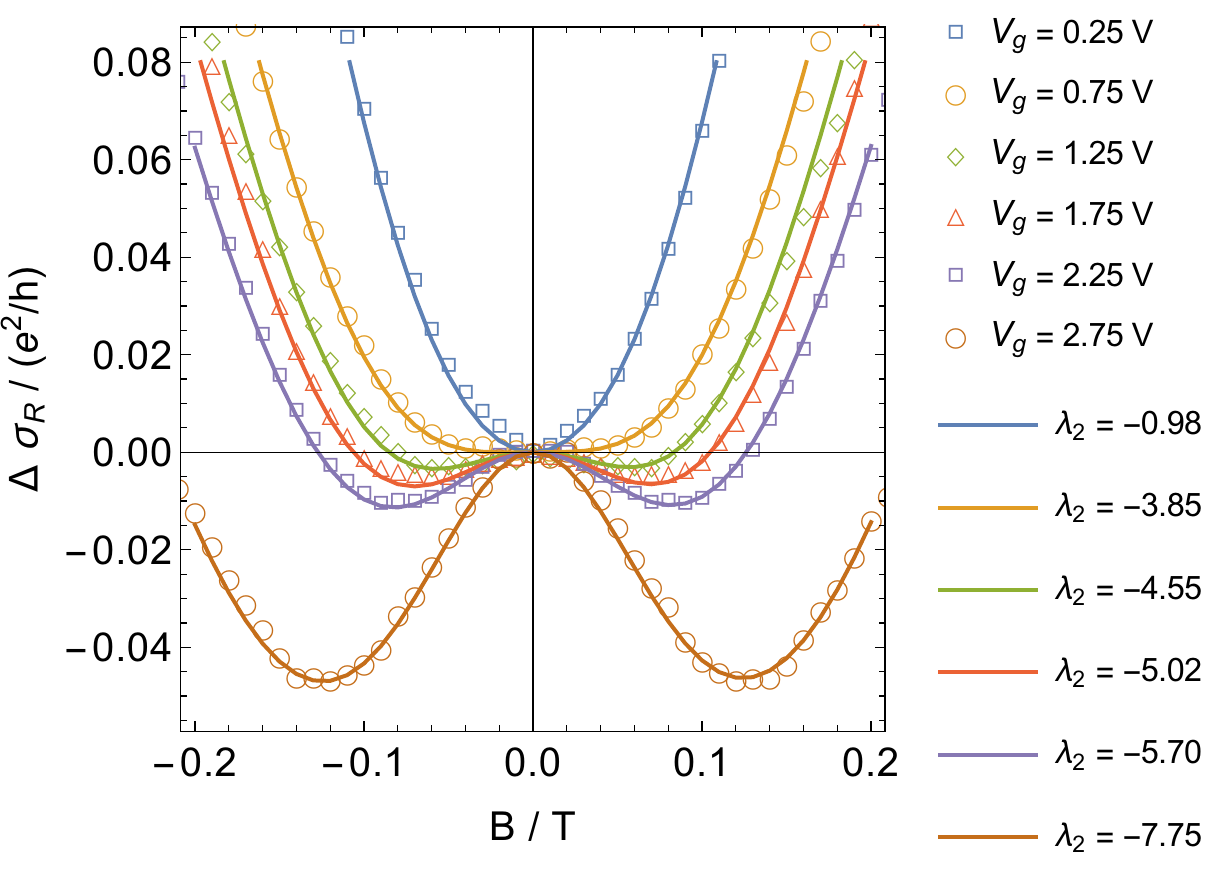}
\caption{(Color online) Gate-controlled crossover from positive to negative magnetoconductivity $\Delta\sigma_R\equiv\Delta\sigma(B)-\Delta\sigma(B=0)$ in a $\braket{111}$ InAs nanowire. The symbol-dotted lines correspond to experimental data for different top-gate voltages $V_g$ which is fitted by theory (solid lines) using Eq.~(\ref{conductivity3}) and varying the external Rashba SOC strength $\alpha_\text{ext}\propto\lambda_2$.}
\label{pic:FitPlotPaper1}
\end{figure}

As an example, we present the fitting results for the magnetoconductance measurements in an undoped top-gated $\braket{111}$ InAs nanowire\footnote{This device corresponds to \textit{Device A} in Ref.~\onlinecite{Heedt2015}.}.
The nanowire is grown by selective area metal-organic vapor phase epitaxy \cite{Akabori2009}. 
Subsequently, the InAs nanowire is transferred to a Si/SiO$_2$ substrate and contacted electrically via electron beam lithography. 
The nanowire segment in-between the source and drain contacts is covered with LaLuO$_3$ high-$k$ dielectric and a metallic gate electrode,~\cite{Heedt2015} giving rise to an external electric field distribution as depicted schematically in Fig.~\ref{pic:wavefct_plot}(a). 
Magnetoresistance measurements are performed in a pumped flow cryostat at a temperature of $1.7\,$K using a low-frequency ($33\,$Hz) lock-in setup with an ac bias current of $10\,$nA. 
As can be seen in Fig.~\ref{pic:FitPlotPaper1}, the device  exhibits a gate-induced crossover from positive to negative magnetoconductivity - which is usually associated with a crossover from WL to WAL.
The same characteristic behavior has been recently observed in several experiments.\cite{Liang2012,Dhara2009,Hansen2005}

The utilization of InAs for nanowires is highly popular\cite{Heedt2015,Dhara2009,Hansen2005,Hernandez2010,
Liang2012,Scheruebl2016}.
In nanowires grown from this material the common  problem of carrier depletion at the surface is avoided as consequence of Fermi level pinning.\cite{Hernandez2010}
The narrow bandgap of InAs results in large Rashba and Dresselhaus SOC coefficients\cite{winklerbook} $r_{41}^{6c6c}=\SI{117.1}{e\angstrom^2}$ and $b_{41}^{6c6c}=\SI{27.18}{eV\angstrom^3}$, respectively.
The effective mass is given by \cite{Bringer2011} $m=0.026 \,m_\text{e}$ where $m_\text{e}$ is the bare electron mass.
In line with the experimental setup of Ref.~\onlinecite{Heedt2015}, we consider a free length of the nanowire of $L=\SI{2.6}{\micro m}$ and a radius of $R_0=\SI{40}{nm}$ where the radial position $R$ of the maximum of the wave function is estimated to be at $R=\SI{35}{nm}$.
Using the relation $\Delta G=(2\pi R/L) \Delta\sigma$ we can determine the conductivity correction from the macroscopic conductance correction $\Delta G$ of the probed nanowire sample.
Moreover, we use the field-effect mobility $\mu=\SI{1000}{cm^2V^{-1}s^{-1}}$ and 3D electron density $n_{3D}=\SI{5.1e17}{cm^{-3}}$ where the 2D electron density can be approximated by $n_\text{2D}=n_\text{3D} R_0^2/(2R)$.
By means of the relation $\mu=e\tau/m$, we find a mean free path of $l_{e}=\SI{17.9}{nm}$ which yields the ratio $l_{e}/(2\pi R_0)=0.08$.
The diffusivity condition around the circumference is, hence, well fulfilled.
Also, the parameter configuration satisfies the Ioffe-Regel criterion with $(\epsilon_F \tau)^{-1}=0.41$ and $\kappa=0.05$ is indeed small.

%

\begin{figure}[t]
\includegraphics[width=.95\columnwidth]{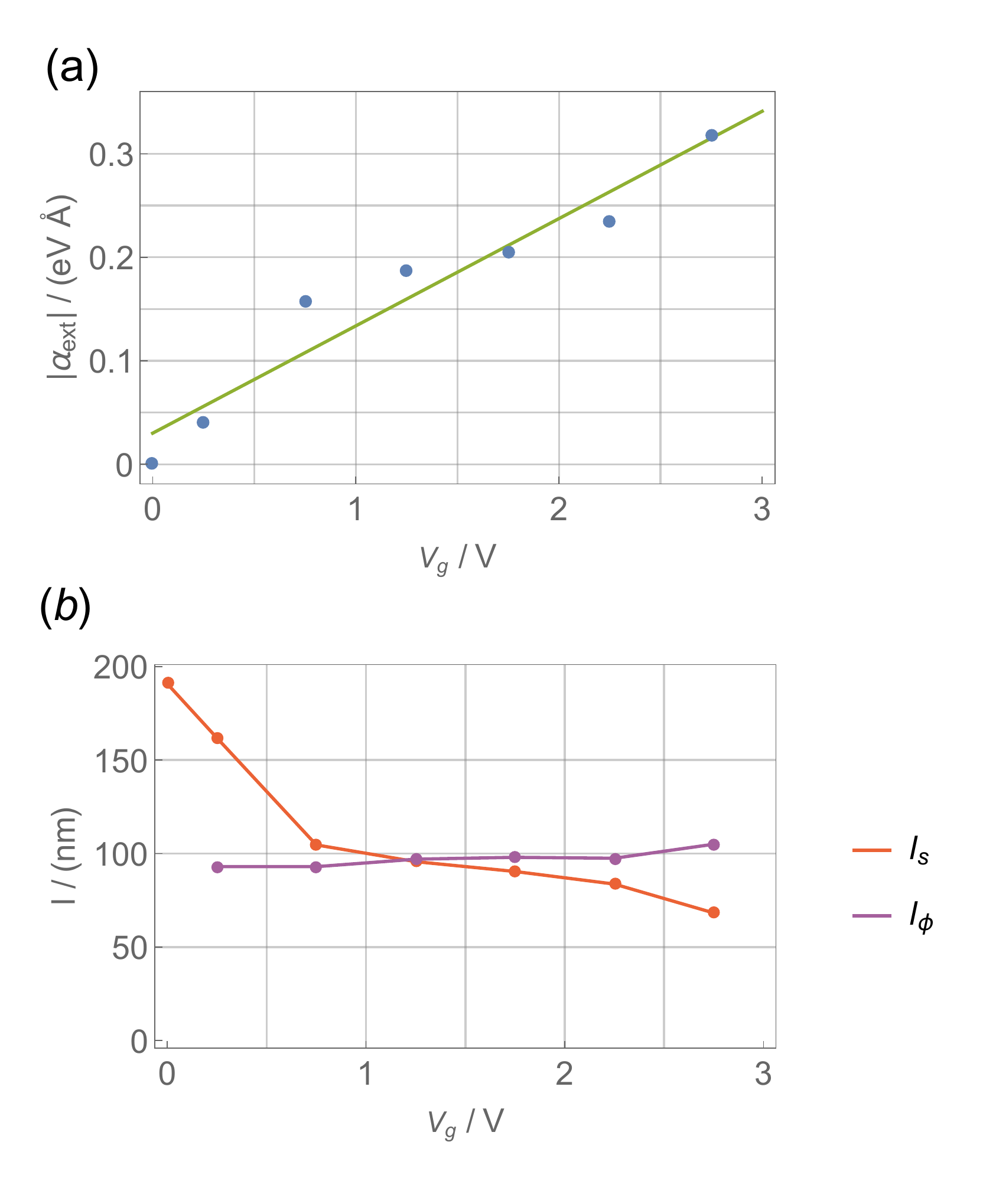}
\caption{(Color online) Extracted fitting parameters for a $\braket{111}$ InAs nanowire. (a) External Rashba SOC strength $|\alpha_\text{ext}|$  as well as (b) spin relaxation and dephasing length $l_s$ and $l_\phi$, respectively, in dependence of a top-gate voltage $V_g$.}
\label{pic:FitPlotPaper2}
\end{figure}

We determine an appropriate fitting value for the internal Rashba contribution of $\alpha_\text{int}=\SI{-74}{meV\angstrom}$.
The respective internal electric field, that arises from Fermi level pinning, is $\mathcal{E}_\text{int}=\SI{-6.3e6}{V/m}$, whose magnitude is in agreement with previous simulations\cite{Heedt2015,Hernandez2010}.
Accordingly, the Dresselhaus SOC strength is found to be $\beta=\SI{41}{meV\angstrom}$ which corresponds to a ratio $\lambda_1=-1.8$ and confinement parameter $\gamma=\SI{0.55}{nm^{-1}}$.
The radial extent of the wave function is pictured in Fig.~\ref{pic:wavefct_plot}(b).
In Fig.~\ref{pic:FitPlotPaper1} we plot the relative magnetoconductivity correction $\Delta\sigma_\text{R}=\Delta\sigma(B)-\Delta\sigma(B=0)$ for an increasing external top-gate voltage $V_g$ or Rashba contribution $\alpha_\text{ext}\propto\lambda_2$, respectively.
A crossover from positive to negative magnetoconductivity due to a growing SOC strength occurs.
The symbol-dotted lines in Fig.~\ref{pic:FitPlotPaper1} illustrate the experimental data, the solid lines the fitted relative magnetoconductivity correction using Eq.~(\ref{conductivity3}).
Each magnetoconductivity curve represents an average of 25 individual measurements in a $\SI{500}{mV}$ gate voltage interval.
In this way, we can ensure that the superimposed universal conductance oscillations are averaged out.
It is shown in Fig.~\ref{pic:FitPlotPaper2}(a), that the scaling between the external Rashba parameter $\vert\alpha_\text{ext}\vert$ and the gate voltage $V_g$ is roughly linear.
The extracted spin relaxation and dephasing lengths $l_s$ and $l_\phi$, respectively, are displayed in Fig.~\ref{pic:FitPlotPaper2}(b) in dependency of an external gate voltage $V_g$.
For a pure internal Rashba contribution, i.e., $\lambda_2=0$, we detect a spin relaxation length of $l_\text{s}=\sqrt{D_{e}\tau_s}=\SI{191}{nm}$ by means of Eqs.~(\ref{spinrelaxtime}) and (\ref{Delta111}).
It decreases simultaneously with an increasing external gate voltage.
In contrast, the dephasing length remains relatively constant at $l_\phi\approx\SI{100}{nm}$.
We stress that, here, we assume the quantum well to remain unchanged as the gate voltage increases.
For high voltages the quantum well width can be expected to become smaller.
However, as consequence of the asymmetry of the external Rashba contribution the associated non-axial symmetric deformation of the quantum well is not comprised in our model for Dresselhaus SOC.


In contrast to previous works, we were able to quantify the Rashba and Dresselhaus SOC parameters individually for a zinc blende type nanowire with surface charge accumulation layer.
The close agreement with experiment in the presented example suggests that the developed model provides reliable information about the transport parameters.

\section{Summary}

Summarizing, we have developed models to describe linear Rashba and Dresselhaus SOC effects in zinc blende semiconductor nanowires of growth directions $\braket{001}$, $\braket{111}$ and $\braket{110}$.
In the considered systems the transport is governed by electron states near the surface which can be a result of Fermi level pinning or radial confinement in core/shell nanowires.
Motivated by recent experiments,\cite{Heedt2015} the Rashba SOC is composed of two parts: an internal and an external contribution.
The internal one is due to an axial symmetric homogeneous electric field induced by Fermi surface pinning or/and a wrap-around gate.
The external one results from an external gate which causes an also axial symmetric but inhomogeneous field that penetrates only one side of the nanowire.
Moreover, we anticipate that the microscopic crystal structure in the nanowire does not differ from the bulk.
This leads to a Dresselhaus SOC which is fundamentally different to previous approaches\cite{Magarill1998,Manulescu2013} that modeled rolled-up $[001]$ confined 2DEGs.
Compared with the latter, the Dresselhaus spin-orbit field depends on the azimuthal location at the surface of the nanowire.

We have computed the Cooperon Hamiltonian following former approaches\cite{Kettemann2007a, PhysRevB.83.115301, Wenk2010, wenkbook}.
The electron motion on the cylindrical surface was treated diffusively in both in-plane coordinates.
It is shown that the Dresselhaus SOC causes a gap for the triplet eigenmodes and, hence, an insuppressible spin relaxation.
This contradicts the conjecture of previous authors, Refs.~\onlinecite{Rainis2014,Dhara2009,Weperen2015,Hansen2005,Nadj2012,
Roulleau2010}, that Dresselhaus SOC  is absent in  $\braket{111}$ nanowires and, hence, cannot cause spin relaxation.
Nevertheless, we found the lowest gap for the $\braket{111}$ growth direction which
indicates a lower spin relaxation than for $\braket{001}$ or $\braket{110}$ nanowires.
A zero-gap mode for certain interplay of Rashba and Dresselhaus SOC which reflects spin-preserving symmetries was not found.
For the $\braket{110}$ nanowires, we observed an additional shift of the minima of the Cooperon modes for the momentum along the wire axis whose value and position depends also on Rashba SOC.
In most cases, it represents the global minimum of the spectrum.
As a consequence, the states with the longest spin lifetime are of helical nature.

Finally, we derived the quantum mechanical correction to the Drude conductivity.
We detected a crossover from negative to positive magnetoconductivity depending on the dephasing time and the SOC strengths.
A significant dependency on the wire radius was not found which was attributed to periodic boundary conditions along the circumference of the cylinder.
By fitting the developed theory to data from low-field magnetoconductance measurements in a $\braket{111}$ InAs nanowire we extracted spin relaxation and dephasing rates as well as SOC strengths.
We were able to quantify the Rashba and Dresselhaus SOC parameters individually.
Both contributions were shown to be likewise significant in a nanowire with surface accumulation layer.

As a final remark, we want to emphasize that studying the magnetoconductance behavior in a nanowire is a particularly delicate task.
The reason is that gating or doping can change the potential landscape or the electron density in such a way that the electron states transform from the surface states (2D) to volume states (3D) in the nanowire.
As the conductivity corrections in 2D and 3D are fundamentally different,\cite{Bruus2002} it is often not clear which model applies.
Additionally, it is ambiguous whether a gate-induced crossover from positive to negative magnetoconductivity is solely attributed to an increase of Rashba SOC or accompanied by a dimensional crossover.
This provides incentive for further studies of the weak (anti-)localization in nanowires where the electron states cover the entire volume.


\section{Acknowledgements}

We gratefully acknowledge Kamil Sladek and Hilde Hardtdegen for nanowire growth and thank Isabel Otto and Thomas Gerster for support with device preparation.
Also, we appreciate the valuable discussions with Klaus Richter, Francisco Mireles, and Andreas Bringer.
This work was supported by Deutsche Forschungsgemeinschaft via Grants No. SFB 689 and No. FOR 912.

%
\appendix
\section{Pauli Matrices in Cylindrical Coordinates}\label{app:pauli}
\begin{align}
\sigma_r={}&
\begin{pmatrix}
0&e^{-i\phi} \\
e^{i\phi}&0
\end{pmatrix},\quad
\sigma_\phi=
\begin{pmatrix}
0&-ie^{-i\phi} \\
ie^{i\phi}&0
\end{pmatrix},\notag\\
\sigma_z={}&
\begin{pmatrix}
1&0 \\
0&-1
\end{pmatrix}.
\label{paulicyl}
\end{align}
%

\section{Commutator Relations}\label{app:com}
\allowdisplaybreaks
\begin{align}
[k_\phi,\cos(\phi)]={}&\,(i/r)\sin(\phi),\\
[k_\phi,\sin(\phi)]={}&\,-(i/r)\cos(\phi),\\
[k_\phi,\sigma_\phi]={}&\,(i/r)\sigma_r,\\
[k_\phi,\sigma_r]={}&\,-(i/r)\sigma_\phi,\\
[k_r,1/r]={}&\,i/r^2,\\
[k_r,k_\phi]={}&\,(i/r)k_\phi.
\end{align}
%

\section{Matrix Elements}\label{app:me}
The matrix elements with respect to the lowest radial mode $\ket{R_0}$ are
\begin{align}
\braket{1/r}={}&1/R,\\
\braket{1/r^2}={}&1/R^2,\\
\braket{k_r}={}&i/(2R),\\
\braket{k_r^2}={}&\gamma^2/2,\\
\braket{k_r^3}={}&3 i \gamma^2/(4 R)\notag\\
={}&3\braket{k_r}\braket{k_r^2},\\
\braket{ 1/r \cdot k_r}={}&0,\\
\braket{ 1/r \cdot k_r^2}={}&\gamma^2/(2 R).
\label{expvhollowcyl}
\end{align}
%
%
\section{Radial Momentum Expectation Value}\label{app:kr}
In this section, we prove that it is not substantial to choose an harmonic radial confinement in order to obtain $\braket{k_r}=i/(2R)$.
A similar proof was demonstrated in Ref.~\onlinecite{Meijer2002}.
Let $\ket{R_0}$ be the lowest radial mode of the Hamiltonian with an arbitrary potential $V(\mathbf{r})$ that confines the wave function $\braket{r|R_0}\equiv\rho_0$ to a region around $R$.
The wave function is demanded to vanish exactly at the limits $r=0$ and $r\rightarrow\infty$. 
We now define $\ket{R_0}\equiv\ket{R_0'}/\sqrt{r}$ and obtain 
\begin{align}
\braket{R_0'|\frac{1}{r}\partial_r|R_0'}=\braket{R_0|\partial_r+\frac{1}{2r}|R_0}=\braket{\partial_r}+\frac{1}{2R}.
\label{krproof1}
\end{align}
On the other hand, partial integration gives 
\begin{align}
\braket{R_0'|\frac{1}{r}\partial_r|R_0'}={}&\int_0^\infty dr (\rho_0')^*\frac{d\rho_0'}{dr}\notag\\
={}&\left|\rho_0'\right|^2\Big|_0^\infty-\int_0^\infty dr \rho_0'\left(\frac{d\rho_0'}{dr}\right)^*.
\label{krproof2}
\end{align}
Since $\left|\rho_0'\right|^2\Big|_0^\infty=r\left|\rho_0\right|^2\Big|_0^\infty=0$, the Eq.~(\ref{krproof2}) must be purely imaginary.
However, given the fact that the Hermiticity of the operator $\frac{1}{r}\partial_r$ requires a real expectation value, $\braket{R_0'|\frac{1}{r}\partial_r|R_0'}$ has to vanish identically.

\section{Correlation Function}\label{app:corr}
The correlation function $\hat{I}$ in Eq.~(\ref{correl}) is evaluated at the Fermi energy $\epsilon_F$. 
For small values of $\mathbf{Q}$ and $1/\tau$ we approximate $\hat{I}$ in the following way. 
For the quasi-2D momentum of the electron we have $\mathbf{q}=(\braket{q_r},q_\phi,q_z)^\top$ and $\ket{\mathbf{q}}=\ket{q_\phi}\ket{q_z}$. 
Defining $(q_\phi,q_z)^\top=(q_\parallel\cos(\vartheta), q_\parallel\sin(\vartheta))^\top$ and the Fermi velocity $\mathbf{v}_F=\mathbf{q}_F/m$ we find with the 2D density of states per spin $\nu=m/(2\pi)$ and the surface of the nanowire $\mathcal{V}$

%
\begin{align}
\hat{I}(\boldsymbol{\hat{Q}})={}&\frac{1}{2\pi\nu\tau\mathcal{V}}\sum_\mathbf{q}\braket{\mathbf{q}|\,\mathcal{G}^\text{R}(\mathbf{\hat{q}},\boldsymbol{\sigma})\mathcal{G}^\text{A}(\mathbf{\hat{Q}}-\mathbf{\hat{q}},\boldsymbol{\sigma'})\,|\mathbf{q}}\notag\\
%
%
%
\approx{}&\frac{1}{\tau m\mathcal{V}}\int_{0}^{2\pi}\frac{d\phi}{2\pi}\sum_{q_\phi,q_z}\frac{1}{\hat{\Sigma}+\frac{i}{\tau}}\frac{\frac{i}{\tau}}{(\epsilon_F-\mathcal{H}(\mathbf{q}))^2+\frac{1}{4\tau^2}}\notag\\
\approx{}&\frac{2\pi i }{\tau m\mathcal{V}}\int_{0}^{2\pi}\frac{d\phi}{2\pi}\sum_{q_\phi,q_z}\frac{1}{\hat{\Sigma}+\frac{i}{\tau}}\delta(\epsilon_F-\mathcal{H}(\mathbf{q}))\notag\\
%
%
\approx{}&\int_{0}^{2\pi}\frac{d\phi}{2\pi}\int_{0}^{2\pi}\frac{d\vartheta}{2\pi}\frac{1}{1-i\tau\hat{\Sigma}}\Bigg|_{\mathbf{q}=m\mathbf{v}_F}
\label{correl2}
\end{align}
%
where $\hat{\Sigma}=\mathcal{H}(\mathbf{\hat{Q}}-m \mathbf{v}_F,\boldsymbol{\sigma})-\mathcal{H}(m \mathbf{v}_F,\boldsymbol{\sigma'})$.

\section{Spin-Orbit Coupling Matrices}\label{app:socmatrices}

The SOC matrix
$\boldsymbol{\hat{a}}=\boldsymbol{\hat{a}}_\text{D}^\xi+\boldsymbol{\hat{a}}_\text{R}$ comprises the Dresselhaus SOC for the different wire directions $\xi \in \left\{[001],[111],[110]\right\}$ as well as internal and external Rashba SOC, i.e.,
$\boldsymbol{\hat{a}}_\text{R}=\boldsymbol{\hat{a}}_\text{R}^\text{int}+\boldsymbol{\hat{a}}_\text{R}^\text{ext}$.
If we chose the basis for convenience in the order
$\{\mathbf{\hat{r}},\boldsymbol{\hat{\phi}},\mathbf{\hat{z}}\}$,
the matrices are written as
\begin{widetext}
\begin{align}
\boldsymbol{\hat{a}}_\text{D}^\text{[001]}&=\,\beta
\begin{pmatrix}
-2\cos(2\phi)&-\frac{5}{2}\sin(2\phi)&0\\
\frac{1}{2}\sin(2\phi)&-\cos(2\phi)&0\\ 
0&0&\cos(2\phi)
\end{pmatrix},\;
\boldsymbol{\hat{a}}_\text{D}^\text{[111]}=\,\frac{\beta}{2\sqrt{3}}
\begin{pmatrix}
0&1&9\sqrt{2} \sin(3\phi) \\
-1&0&3\sqrt{2} \cos(3\phi)\\ 
-\sqrt{2} \sin(3\phi)&-\sqrt{2} \cos(3\phi)&0
\end{pmatrix},\notag\\
\boldsymbol{\hat{a}}_\text{D}^\text{[110]}&=\,\frac{\beta}{8}
\begin{pmatrix}
0&0&-\cos(\phi)-27\cos(3\phi) \\
0&0&\sin(\phi)+9\sin(3\phi)\\ 
\cos(\phi)+3\cos(3\phi)&-11\sin(\phi)-3\sin(3\phi)&0
\end{pmatrix},\\
\intertext{and}
\boldsymbol{\hat{a}}_\text{R}^\text{int}&=\,\alpha_\text{int}
\begin{pmatrix}
0&0&0 \\
0&0&-1\\ 
0&1&0
\end{pmatrix},\;
\boldsymbol{\hat{a}}_\text{R}^\text{ext}=\,\alpha_\text{ext}\Theta(\phi)\Theta(\pi-\phi)
\begin{pmatrix}
0&0&\cos(\phi) \\
0&0&-\sin(\phi)\\ 
0&\sin(\phi)&0
\end{pmatrix}.
\label{dressmatrix}
\end{align}
\end{widetext}
%
%
%

%

\section{Singlet-Triplet Representation and Sum Formula}\label{app:st}
The $\ket{s_1,s_2}$ basis of the spin $z$-components of the two
electrons with $s_i \in \{+,-\}$, labeled by $(\pm)$, can be
transformed into the singlet-triplet representation $\ket{s,m_s}$ with
$s \in \{0,1\}$ and $m_s \in \{0,\pm1\}$ by the relations
\begin{align}
\ket{0,0}={}&\frac{1}{\sqrt{2}}(\ket{+,-}-\ket{-,+}),\\
\ket{1,0}={}&\frac{1}{\sqrt{2}}(\ket{+,-}+\ket{-,+}),\\
\ket{1,\pm1}={}&\ket{\pm,\pm}.
\end{align}
This yields the unitary transformation matrix
\begin{equation}
U=\,\frac{1}{\sqrt{2}}
\begin{pmatrix}
0&\sqrt{2}&0&0\\
-1&0&1&0\\
1&0&1&0\\
0&0&0&\sqrt{2}
\end{pmatrix}.
\end{equation}

Hence, the spin matrices in singlet-triplet representation become
$\hat{S}_i=U^\dag \hat{S}_i^{(\pm)} U$, or particularly
\begin{align}
S_x={}&
\,\frac{1}{\sqrt{2}}
\begin{pmatrix}
0&0&0&0\\
0&0&1&0\\
0&1&0&1\\
0&0&1&0
\end{pmatrix},\notag\\
S_y={}&
\,\frac{i}{\sqrt{2}}
\begin{pmatrix}
0&0&0&0\\
0&0&-1&0\\
0&1&0&-1\\
0&0&1&0
\end{pmatrix},\notag\\
S_z={}&
\,
\begin{pmatrix}
0&0&0&0\\
0&1&0&0\\
0&0&0&0\\
0&0&0&-1
\end{pmatrix}
\label{spinst}
\end{align}
in the order $\{\ket{0,0},\ket{1,1},\ket{1,0},\ket{1,-1}\}$. Thus, the
singlet and triplet sector decouple.

In singlet-triplet representation 
the sum over spin indices $s_1,s_2$ in
Eq.~(\ref{conductivity}) simplifies to
\begin{align}
\sum_{s_1,s_2=\pm}&\braket{s_1,s_2|\hat{\mathcal{C}}|s_2,s_1}=\notag\\
&=\text{Tr}\left[\Lambda \hat{\mathcal{C}}^{(\pm)}\right]=\text{Tr}\left[\Lambda U \hat{\mathcal{C}}U^\dag\right]=\text{Tr}\left[U^\dag\Lambda U \hat{\mathcal{C}}\right]\notag\\
&=-\braket{0,0|\hat{\mathcal{C}}|0,0}+\sum_{m_s}\braket{1,m_s|\hat{\mathcal{C}}|1,m_s}\notag\\
&=\frac{1}{D_e}\left(-\frac{1}{E^\mathcal{S}}+\sum_{i}\frac{1}{E^\mathcal{T}_i}\right),
\label{sumformula1}
\end{align}
where
\begin{equation}
\Lambda=
\begin{pmatrix}
1&0&0&0\\
0&0&1&0\\
0&1&0&0\\
0&0&0&1
\end{pmatrix}
\end{equation}
and thus
\begin{equation}
U^\dag\Lambda U=
\begin{pmatrix}
-1&0&0&0\\
0&1&0&0\\
0&0&1&0\\
0&0&0&1
\end{pmatrix}.
\end{equation}
\\

\newpage
\bibliographystyle{apsrev4-1}
\bibliography{WK}
\end{document}